\documentstyle[preprint,tighten,aps,epsfig]{revtex}
\begin{document}

\title{Field Theory of the Random Flux Model}

\author{Alexander Altland$^{1)}$ and B D Simons$^{2)}$}
\address{1) Theoretische Physik III, Ruhr-Universit\"at Bochum, 44780 Bochum,
   Germany,\\ 
   2) Cavendish Laboratory, Madingley Road, Cambridge CB3\ OHE,
  UK}

\date{\today}
\maketitle 
\draft  
\begin{abstract} 
  The random flux model (defined here as a model of lattice fermions
  hopping under the influence of maximally random {\it link} disorder)
  is analysed field theoretically. It is shown that the long range
  physics of the model is described by the supersymmetric version of a
  field theory that has been derived earlier in connection with
  lattice fermions subject to weak random hopping. More precisely, the
  field theory relevant for the behaviour of $n$-point correlation
  functions is of non-linear $\sigma$ model type, where the group
  ${\rm GL}(n|n)$ is the global invariant manifold. It is argued that
  the model universally describes the long range physics of random
  phase fermions and provides further evidence in favour of the
  existence of delocalised states in the middle of the band in two
  dimensions. The same formalism is applied to
  the study of non-Abelian generalisations of the random flux model,
  i.e. $N$-component fermions whose hopping is mediated by random
  ${\rm U}(N)$ matrices. We discuss some physical applications of
  these models and argue that, for sufficiently large $N$, the
  existence of long range correlations in the band center (equivalent
  to metallic behaviour in the Abelian case) can be safely deduced
  from the RG analysis of the model.
\end{abstract} 


\section{Introduction}
\begin{figure}[b]
\begin{center} 
\epsfig{file=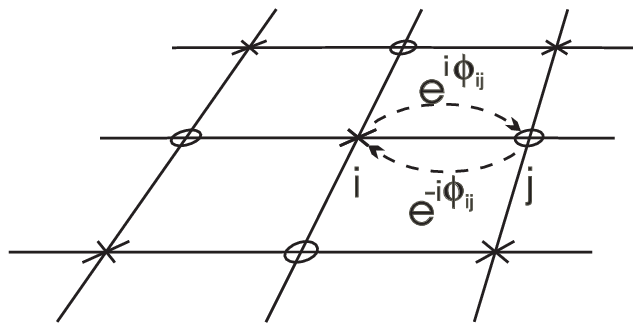,height=4cm}
\end{center}
\caption{The two-dimensional random flux model.}
\label{fig:rf}
\end{figure}

Quantum disordered systems are typically described in terms of 
Hamiltonians of the form
\begin{equation}
\hat H = \hat H_0 + \hat V,
\end{equation}
where $\hat H_0$ models the underlying clean system and
disorder is introduced via the randomly distributed {\it Hermitian}
operator $\hat V$. Sometimes, however, it is preferable to describe
disordered systems differently, viz. by Hamiltonians where disorder is
implemented in terms of {\it unitary} stochastic operators. 
A simple
prototype example is the system of lattice electrons subject to a
(perpendicular) strong random magnetic flux (see fig.~\ref{fig:rf}).
An electron hopping from one site to the next acquires a field
dependent phase which, if the field is strong
enough\footnote{More precisely, maximum
  stochasticity requires fields stronger than one flux-quantum per
  lattice cell.} becomes completely random. Under these conditions the
system is described by the Hamiltonian,
\begin{equation}
  \label{H_rf}
  \hat H=-\sum_{\langle i, j\rangle} c_i^\dagger e^{i\phi_{ij}} c_j,
\end{equation}
where $\sum_{\langle i,j\rangle}$ is a summation over all nearest
neighbour sites of a $d$-dimensional bipartite lattice, $c_i$ is an
electron field operator at site $i$, $\phi_{ij}$ a phase living on the
link $i\to j$, and maximum stochasticity implies that the phase
$\phi_{ij}$ is drawn randomly from the full interval $[0,2\pi]$
subject only to the Hermiticity requirement $\phi_{ij}=-\phi_{ji}$.
Notice that the definition (\ref{H_rf}) implicitly assumes that all
energies are measured in units of the modulus of the hopping matrix
elements.

It has been suggested that the model (\ref{H_rf}) describes the physics of
the fractional quantum Hall effect close to half filling (where the
problem is expected to map onto a system of composite fermions subject
to a strong random field with zero mean~\cite{Halperin}). 
Furthermore, it is believed to be of some relevance in the gauge theory 
of high-$T_c$ superconductivity~\cite{Nagaosa}. (However, it is important to 
recognise that the denotation `random flux model' is, to some extent, 
misleading in the sense that the Hamiltonian (\ref{H_rf}) does {\it not} 
actually model a short range correlated random magnetic field
but rather the lattice version of a random {\it vector potential}. The
difference can be of importance since a short range correlated field
would imply the presence of {\it long} range correlations in the
vector potential~\cite{Gavazzi,Aronov}. Keeping this difference in 
mind we nevertheless continue to refer to (\ref{H_rf}) as a random flux
Hamiltonian.)

\begin{figure}[hbt]
\begin{center} 
\epsfig{file=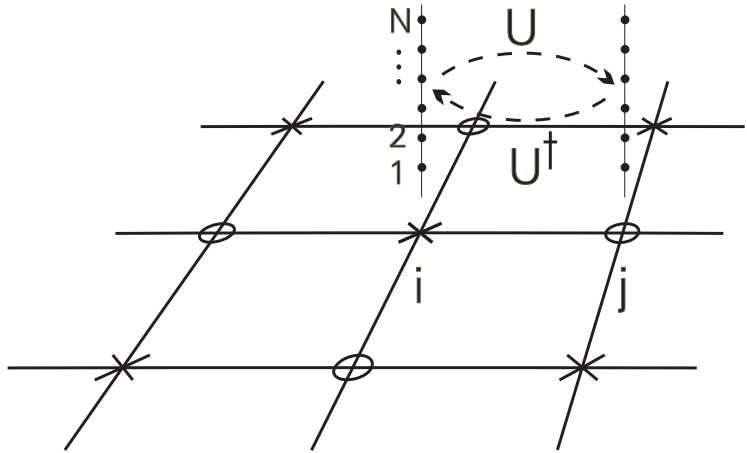,height=6cm}
\end{center}
\caption{The generalized random flux model.}
\label{fig:UN}
\end{figure}

The random phase  Hamiltonian (\ref{H_rf}),
describing the lattice dynamics of {\it scalar} fermions, can
straightforwardly be generalized so as to accommodate fermions with an
internal structure: Let us assume that a) the lattice fermions carry
an internal quantum number, $c_i \to c_i^\alpha, \alpha = 1,\dots,N$
such as, e.g., spin and b) that the fundamental hopping mechanism
leads to a full randomisation of the internal states. Under these
circumstances, the Hamiltonian (\ref{H_rf}) generalises to (c.f.
fig.~\ref{fig:UN})
\begin{equation}
  \label{UN_H}
  H=-\sum_{\langle i, j \rangle} c_i^\dagger U_{ij} c_j,
\end{equation}
where $U_{ij}\equiv\{ U_{ij}^{\alpha,\beta} \} \in {\rm U}(N)$ are 
$N$-dimensional unitary matrices sitting on the nearest
neighbour links of the lattice, and $c_i \equiv \{c_{i,a}^\alpha
\}$, $\alpha=1,\dots,N$, $a=1,\dots,n$ are $N\times n$-component field
operators. The lower index $a$, which has been introduced for future
reference, plays a passive role in the sense that it is summed over
and does not couple to the $U$'s. `Full randomisation' upon each
hopping process translates to the $U_{ij}$ being drawn
democratically from the Haar measure of the group (subject to the
Hermiticity requirement $U_{ij}=U_{ji}^\dagger$.) Following a convention 
introduced in Ref.~\cite{Zirnbauerb} we refer to the index
$\alpha$ as a `colour' index, whilst $a$ will be denoted as a
`flavour' index (for the motivation of this terminology, see below).
The original RF-Hamiltonian fits into the general framework in the
sense that it is the $N=1$ version of (\ref{UN_H}). For the sake of
simplicity we will also refer to (\ref{UN_H}) as a `random flux'
model.

The generalized RF-model has applications in both condensed matter and
high energy physics:

{\it Physics of manganese oxides:} Suggestions have been put forward
to explain the effect of giant magnetoresistance, shown by transition
metal oxide compounds, in terms of effective models consisting of
lattice electrons hopping in a matrix of localised $S=3/2$ core
spins\cite{Anderson}. The hopping process is essentially influenced by
the Hund's rule coupling between the spin of the itinerant electrons
and the localised spins. In the limit where the coupling strength
becomes infinite, the hopping matrix element along any given bond
$i\to j$ depends solely on the orientation of the spins at the sites
$i$ and $j$\cite{MuHa}, i.e. the hopping is mediated by an ${\rm
  SU}(2)$ matrix which acts in the $S=1/2$ space of the electron spins
and depends on the orientation of the two adjacent core spins.
Assuming that a) the spin orientations are random, b) that the effect
of the randomness can be mimicked by drawing the {\it link} matrices
from a random distribution, and c) that, with regard to universal
properties, the difference between random ${\rm SU}(2)$ hopping and
random ${\rm U}(2)$ hopping is of no concern, we arrive at the
$(N=2,n=1)$ version of (\ref{UN_H}).

{\it Lattice gauge theory:} The simplest\footnote{A more realistic
  lattice representation of the QCD Lagrangian would involve Dirac
  $\gamma$-matrices acting in the Dirac-spinor space of the fermionic
  matter field. Although the embedding of this additional structure
  into the formalism discussed below is straightforward, we prefer to
  keep the discussion simple and not to include the Dirac-spinor
  structure.} lattice action of a matter field with $N$ `colour'
indices subject to a non-Abelian local gauge symmetry is given by
\begin{equation}
  \label{Slatt}
  S=-\sum_{\langle i, j \rangle} 
c^{a\dagger}_i U_{ij} c^a_j -
\frac{1}{2g}\sum_\Box \left[\left(1-\frac{1}{N}{\rm tr}_\Box(U^4)\right) +
{\rm h.c.}\right].
\end{equation}
Here, $a=1,\dots,n$ has the physical significance of a fermionic
flavour index, $\sum_\Box$ is a sum over all plaquettes of the
lattice, and ${\rm tr}_\Box(U^4) \equiv {\,\rm
  tr\,}\left(U_{ij}U_{jk}U_{kl}U_{li}\right)$ represents the trace
over all $U$'s sitting on the links of the plaquette. Physically, the
plaquette action represents the discrete version of a continuum field
strength tensor (c.f., for example,~\cite{creutz}). In the strong coupling
limit, $g\to \infty$ (i.e. the long distance-confinement phase of the
theory) the gauge fields become free and (\ref{Slatt}) is governed by
a structure like (\ref{UN_H}).

What can be said, in general terms, about the physical behaviour of
the scalar random flux Hamiltonian and its non-abelian generalizations?:
\begin{itemize}
\item At first sight it looks like the random flux Hamiltonian falls
  into the general category of Anderson Hamiltonians for disordered
  systems. More precisely, the model (\ref{H_rf}) bears similarity
  with a lattice Anderson Hamiltonian with strong disorder in the
  hopping matrix elements. Being Hermitian but not real, the
  Hamiltonian (\ref{H_rf}) is expected to fall into the general class
  of models with unitary symmetry. In particular, all states might be
  expected to be Anderson localised. As will be discussed below, this
  picture is correct as long as one stays away from zero energy:
\item {\it The middle of the tight binding band}, $\epsilon=0$. In the
  band center, the Hamiltonian (\ref{H_rf}) possesses a discrete
  symmetry which, as will be explored in detail below, heavily
  influences the physical properties of the model. The nature of the
  symmetry can easily be understood by inspection of fig.~\ref{fig:rf}: 
  The essential point is that the Hamiltonian of the RF
  problem is of pure nearest neighbour type. (There are no matrix
  elements connecting any one site with itself.) To connect this
  statement with the existence of a symmetry, one may subdivide the
  lattice into two sublattices (the $\circ$'s and the $\times$'s in
  fig.~\ref{fig:rf}). When represented in a  $\circ/\times$ block
  decomposition the Hamiltonian becomes purely off-diagonal,
  \begin{equation}
    \label{chiral1}
    \hat H = \left(
      \begin{array}{cc}
      &\hat H_{\circ,\times}\\
      \hat H_{\times,\circ}&
      \end{array}\right).
  \end{equation}
  Formally, the block `off-diagonality' of the the Hamiltonian can be
  expressed as
  \begin{equation}
    \label{chiral2}
    \sigma_3 \hat H \sigma_3 = - \hat H \Leftrightarrow [\hat H,
    \sigma_3]_+=0,
  \end{equation}
  where $[\;,\;]_+$ is the anticommutator and $\sigma_3$ a Pauli
  matrix in $\circ/\times$ space. Symmetries of the type
  (\ref{chiral2}) are commonly referred to as {\it chiral
  symmetries}. 

The existence of a chiral symmetry has far reaching consequences for
the physics of the model at the band-center. Before turning to a more
comprehensive discussion of this point, let us summarise a few
elementary facts that follow straightforwardly from the definition of
the model.  Firstly, we note that the Hamiltonian is {\it always}
chiral, i.e. not just in the band center. One may then ask why, for
finite energies, the chiral symmetry is ineffective. The reason is
that the information about both thermodynamic and transport properties
of the system is encoded in the single-particle Green function
$G^\pm(\epsilon) \equiv (\epsilon^\pm - \hat H)^{-1}$ rather than in the
Hamiltonian itself (here $\epsilon^\pm \equiv \epsilon \pm i 0$). Now,
\[
  G^\pm(\epsilon) = -\sigma_3 G^\mp(-\epsilon) \sigma_3,
\]
i.e. the inversion property under adjungation with $\sigma_3$ is
broken by finite $\epsilon$. Some physical implications of this
symmetry breaking will be discussed in detail below. 

Secondly, we note that stochastic Hamiltonians with a chiral structure have 
previously been studied in the random matrix
literature~\cite{v1,v2,v3,ns,hikami1,hikami2}. It has been shown that
they belong to symmetry classes that are fundamentally different from
the standard Wigner Dyson classes. In particular, the band center RF
model belongs to the universality class of the
so-called `chiral GUE', ChGUE~\cite{ast,v1,v2,v3,ns,hikami2}. For
finite $\epsilon$, the model undergoes a symmetry crossover from ChGUE
to ordinary GUE symmetry.

To the best of our knowledge the full extent of the consequences of the 
chirality of the RF model was first noticed in a seminal paper by Miller and 
Wang~\cite{Miller}.

\item Beginning with Lee and Fisher~\cite{Lee81}, there has been intensive 
  research into the Anderson localization behaviour of the model. So far 
  the question whether or not all states are localised has not been 
  conclusively settled: Despite extensive numerical and analytic studies,
  opinion is divided between those who conclude that all states are 
  localised~\cite{Sugiyama,Lee,Aronov,Kim,Yakubo,Batsch}, and those 
  who believe that delocalised states persist at the center of the 
  band~\cite{Pryor,Avishai,Kalmeyer,Zhang,Liu,Sheng,Miller,Yang,Xie,Furasaki}.
  Below we will provide further 
  evidence for the existence of a metallic phase in the middle of the
  band. Needless to say, the phenomenon is induced by the
  unconventional symmetry of the model. By contrast, for finite
  $\epsilon$, the model can be shown to map onto the standard Anderson
  model of unitary symmetry which is known to be localised in
  dimensions two and less. For weak random magnetic fields, this
  connection has already been established microscopically in 
  Ref.~\cite{Aronov}.

  Besides the localization behaviour, several other physical properties
  of the RF system are affected by the chiral symmetry. In
  particular the presence of the symmetry gives rise to non-trivial
  quantum-interference corrections to average {\it single}-particle 
  properties such as, e.g. the density of states.  Some of
  these features will be discussed below. 
\end{itemize}

The subject of this paper is the construction of an effective field
theory describing the low energy/long range properties of the
generalized random flux model (\ref{UN_H}). Some emphasis will be placed
on exploring the consequences of the chiral symmetry of these systems.
Specifically, it will be shown that the random flux model maps onto a
field theory of a structure similar to that constructed earlier by
Gade~\cite{gade} in connection with {\it weakly} disordered sublattice
models introduced by Oppermann and Wegner\cite{wegner}. Both the symmetries 
and the
structure of Gade's field theory are known to be essentially
influenced by chirality. In particular, renormalization group (RG)
flow of the theory is drastically different from the one
observed in conventional low dimensional disordered systems. Indeed, the fact
that the RF models (\ref{UN_H}) and the Wegner/Oppermann sublattice systems of
broken time reversal invariance, respectively, map onto the same field
theory might have been anticipated from the outset, the
reason being that the latter represents the maximally 
disordered counterpart of the former. On the field theoretical side,
the main {\it difference} between the models lies in the value of the
coupling constants. The consequences of this difference for the large
scale analysis of the RF model will be discussed below, after the
field theory has been microscopically derived. A second difference to
Gade's analysis is that the present construction will be based on
the supersymmetry formalism (as opposed to Gade's boson replicas). As
will be discussed below, the supersymmetric formulation enables one to
explicitly demonstrate that the ergodic limit of the model is
described by chiral random matrix theory.

The paper is organized as follows: We start out from a qualitative 
explanation for the importance of the chiral symmetry in section
\ref{sec:qualitative}. The central sections \ref{sec:cf} and
\ref{sec:effective} contain the derivation of the long range field
theory of the model. In section \ref{sec:eval} we discuss what kind of
physical predictions follow from the field theoretical modelling. 
To clarify our discussion, much of the technical content of the analysis has
been moved to appendices.

\section{The Chiral Symmetry: Qualitative Considerations}
\label{sec:qualitative}
Prior to embarking on the quantitative analysis of the RF model, it is
instructive to develop some intuition regarding the consequences of
the presence of the chiral symmetry. Perhaps the most apparent
manifestation of the symmetry is the non-triviality of ensemble averaged 
{\it single-particle} properties such as the DoS: As shown in Ref.~\cite{gade}
within a perturbative renormalization group scheme, the DoS of
the two-dimensional (weakly disordered) sublattice model diverges upon
approaching the middle of the band. The same type of behaviour has
been predicted for one-dimensional chiral systems~\cite{Theodorou}. 
That these results can, in fact, not hold true all the way down to
$\epsilon=0$ follows from exact random matrix 
theory~\cite{v1}. The random matrix analysis which, by general
symmetry arguments, is expected to effectively describe the ergodic
limit of chiral problems (i.e. the asymptotic
vicinity of the band center) yields a {\it vanishing} DOS at
$\epsilon=0$. Since the breakdown of the DoS takes place on an energy
scale set by the level spacing, it is generally difficult to detect
within perturbative approaches.  In fact it is known, for example from the
conceptually related analysis of the level-spacing gap in the DoS of
superconductor/normal metal systems without proximity effect~\cite{az},
that a non-perturbative level spacing gap may signal its presence in
terms of spurious {\it divergences} appearing in perturbative
analyses. Thus, judging from perturbation theory alone, it may be
difficult to tell what part of the band center DoS divergence is
`real' and what is merely an unphysical repercussion of a tiny
non-perturbative effect.

As will be discussed in more detail below, the complex behaviour of
the DoS finds its origin in the fact that, in chiral models, mechanisms of 
large scale quantum interference --- which are commonly represented by
diffusion-type modes like `diffusons' and `Cooperons' --- already
operate on the level of single-particle properties. This should be
contrasted with the situation in conventional disordered models where
the average Green function picks up a largely structureless imaginary
self energy but otherwise is not affected by the disorder.

\begin{figure}
\begin{center} 
\epsfig{file=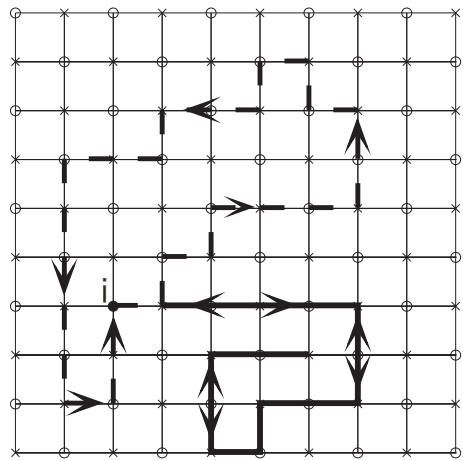,height=6cm}
\end{center}
\caption{\label{fig:self_ret} Dashed: generic path contributing to
  (\protect\ref{DoS_path}). Solid: Self retracing path which survives
  phase averaging.}
\end{figure}

To understand why the DoS of the RF model behaves unusually in the
band center, it is not even necessary to invoke the concept of
diffusion modes. A simple path counting argument analogous to that employed
by Brinkman and Rice~\cite{Brinkman} is sufficient:
Consider the average DoS per unit volume, $\rho$, represented in terms of
the single-particle Green function,
\begin{equation}
  \label{DoS_def}
  \rho(\epsilon) = \frac{1}{\pi} {\,\rm Im\,} \sum_i \left\langle 
  \left\langle i\left|
  \frac{1}{\epsilon^--\hat H }\right|i\right\rangle \right\rangle,
\end{equation}
where the outer brackets represent the phase averaging
\begin{equation}
\label{phase_av}
\left\langle \dots \right \rangle \equiv \prod_{\langle
  i,j\rangle} \frac{1}{2\pi} \int_0^{2\pi}d\phi_{ij} (\dots).
\end{equation}
Expanding in powers of $\hat H$ and inserting lattice resolutions of
unity, $\sum_j |j\rangle \langle j|$, we obtain
\begin{eqnarray}
  \label{DoS_path}
&&\rho(\epsilon) = -\frac{1}{\pi}  {\,\rm Im\,} \sum_{n=0}^\infty
\sum_{i,i_1,\dots,i_n} (-\epsilon^-)^{-(n+1)} \left.\left\langle
e^{i\sum_{m=0}^{n-1} \phi_{i_m,i_m+1}}\right\rangle\right|_{i_0 =i_n =
i}\nonumber\\&&\qquad= 
\frac{1}{\pi}\left(\frac{L}{a}\right)^d  {\,\rm Im\,} \sum_{n {\,\rm even\,}}^\infty
\sum_{i_1,\dots,i_n \atop {\rm s.r.}} (\epsilon^-)^{-(n+1)}\cdot 1 \nonumber\\
&&\qquad=\frac{1}{\pi}\left(\frac{L}{a}\right)^d   {\,\rm Im\,}
\sum_{n {\,\rm even\,}}(\epsilon^-)^{-(n+1)} (2d)^{n/2}, 
\end{eqnarray}
where $L$ is the linear extension of the lattice and $a$ the lattice
spacing.  The expression $\sum_{m=0}^{n-1} \phi_{i_m,i_m+1}$ has the
meaning of the total phase acquired during the propagation along the
path $i\to i_1\to i_2 \to \dots \to i$ (c.f. fig.~\ref{fig:self_ret}).
The second equality in (\ref{DoS_path}) states that the
contribution of all paths with a non-vanishing total phase averages to
zero. The surviving $0$-phase paths are self retracing --- indicated
by the symbol `s.r.'.  In the third equality we have used the fact
that the total number of self retracing paths of length $n$ is
$(2d)^{n/2}$. Evidently, the summation over $n$ in the last line
diverges for $\epsilon<\sqrt {2d}$, indicating a breakdown of
perturbation theory in the vicinity of the band center.  Notice that
the presence of $0$-phase paths is intimately connected to the
sublattice structure of the system. No analogous contribution to the
mean DoS of conventional (on-site) disordered systems exists.

A second and more formal way of demonstrating the non-triviality of
single-particle properties is based on a few algebraic manipulations
of the single-particle Green function. Introducing a system of Pauli
matrices in the two-component $\circ/\times$-partition, the
Hamiltonian takes the form
\[
\hat H = \left(
  \begin{array}{cc}
0 &\hat H_1 - i \hat H_2\\
\hat H_1 + i \hat H_2 & 0
  \end{array}\right)=U^\dagger\left(
  \begin{array}{cc}
\hat H_1& - i \hat H_2\\
i \hat H_2 & -\hat H_1
  \end{array}\right) U,
\]
where $U=\exp(i\frac{\pi}{4}\sigma_2)$ and both $\hat H_1$ and $\hat
H_2$ are Hermitian. Similarly, the Green function can be represented
as
\[
G^+(\epsilon) = U^\dagger\left(  \begin{array}{cc}
\epsilon^+ -\hat H_1& + i \hat H_2\\
-i \hat H_2 & \epsilon^+ + \hat H_1
  \end{array} \right)^{-1}U= U^\dagger\left(  \begin{array}{cc}
\epsilon^+ - \hat H_1&   i\hat H_2\\
i \hat H_2 & -\epsilon^+ - \hat H_1
  \end{array} \right)^{-1}\sigma_3 U .
\]
The point now is that the central matrix can be interpreted as a
retarded and an advanced stochastic Green function
$g^+(\epsilon)\equiv (\epsilon^+ - \hat H_1)^{-1}$ and
$g^-(-\epsilon)\equiv (-\epsilon^+ - \hat H_1)^{-1}$, coupled by the
operator $i \hat H_2$. Since the the manipulations with the matrices
$U$ and $\sigma_3$ do not affect the structure of the singularities of
the Green function, we conclude that the single-particle
$G^+(\epsilon)$ has the effective complexity of {\it two} coupled
stochastic Green functions. In particular, we can anticipate that the
average $G^+(\epsilon)$ will be affected by mechanisms of quantum
interference between $g^+(\epsilon)$ and $g^-(-\epsilon)$. Notice that
both large energy arguments $\epsilon$ and on-site matrix elements of
the Hamiltonian (corresponding to diagonal blocks in the
$\circ/\times$-partition) would lead to a destruction of this
two-particle coherence.

After these general considerations, we proceed to the quantitative study
of the generalized RF-problem. Our general strategy parallels that employed
by Zirnbauer~\cite{Zirnbauerc} in a loosely related study of the mapping
of the Chalker-Coddingtom network model of the Integer Quantum Hall Effect
onto the effective field theory proposed by Pruisken.

\section{Mapping the Random Flux Model to a Field Theory:
Colour-Flavour Transformation}
\label{sec:cf}
As a first step towards the construction of a field theory of the RF
problem, we need to introduce some degrees of freedom capable of
describing the long range features of the model (\ref{UN_H}).  A
mathematical tool for mapping model systems governed by unitary
stochastic operators onto functional integrals containing a `low
energy' sector describing the large distance physics has recently been
introduced and applied by Zirnbauer~\cite{Zirnbauerb}. For reasons that
will become clear below, the transformation has been christened the
`colour-flavour transformation'. Our application of the colour-flavour
transformation to the ${\rm U}(N)$ RF-model closely parallels the
analysis of Ref.~\cite{Zirnbauerb} of the spectral properties of {\it
  single} ${\rm U}(N)$ matrices. For this reason the presentation of
the initial stages of the construction of the theory will be
comparatively concise but nevertheless self contained. (For a deeper
discussion of the mathematics of the colour-flavour transformation we
refer the reader to the original reference.)

As a first step towards the transformed theory, we consider the
Gaussian functional integral
\begin{equation}
  \label{Z1}
Z[\hat J]=\left \langle \int {\cal D}(\psi,\overline\psi)\;
e^{\hspace{0.3cm}i\hspace{-0.3cm}\sum\limits_{\langle i\in A, j\in B
    \rangle}\left\{ 
\overline\psi_i s^{1/2}  U_{ij} s^{1/2} \psi_j +
\overline\psi_j s^{1/2} U_{ij}^\dagger s^{1/2} \psi_i \right\}
+i\sum\limits_{k,l} \overline\psi_k s^{1/2}
 \left(\hat{z}\delta_{kl} +\hat J_{kl}\right) s^{1/2}
 \psi_l}\right\rangle,  
\end{equation}
where 
\[
\left\langle \dots \right \rangle \equiv \prod_{\langle
  i,j\rangle} \int dU_{ij} (\dots)
\]
is the non-Abelian generalisation of the phase average
(\ref{phase_av}), $dU_{ij}$ denotes the Haar measure on ${\rm U}(N)$
and $A$ ($B $) is the sublattice of $\circ$-sites ( $\times$-sites).
The fields
\[
\bar \psi = \{\bar \psi_{i,a}^\alpha \}, \qquad  \psi = \{\psi_{i,a}^\alpha \}
\]
carry three types of indices: A lattice index $i$, a `colour' index
$\alpha$ (coupling to the internal group indices of the $U$'s), and a
two component `flavour' index $a\equiv(\sigma,m)$. Here $m=1,\dots,n$
is a discrete counting variable whose significance will become clear
shortly, whereas $\sigma=1,2$ introduces a boson/fermion grading into
the theory: Field components $\psi^\alpha_{i,\sigma =1,m}\equiv
S^\alpha_{i, m}$ are complex commuting whilst $\psi^\alpha_{i,\sigma
  =2,m}\equiv \chi^\alpha_{i,m}$ are odd elements of a Grassmann
algebra (`anticommuting variables'). Convergence of the integral
(\ref{Z1}) requires that the components $\bar S^\alpha_{i, m}$ be the
complex conjugates of $S^\alpha_{i, m}$, while the variables 
$\chi^\alpha_{i,m}$ and $\bar \chi^\alpha_{i,m}$ are independent of each 
other. In cases where
no confusion can arise we will employ the more explicit notation,
$\sigma={\rm B, F}$ instead of $\sigma=1,2$.

The elements of the diagonal matrix 
\[
\hat z = {\rm diag}(z_1,\dots,z_n)\otimes
\openone_{2N},
\]
where (see below) $z_m \in {\cal C}$, have the significance of energy 
arguments of single-particle Green functions. The presence of the matrix
\[
s={\rm diag}({\rm
  sgn\, Im\,}z_1,\dots,{\rm
  sgn\, Im\,}z_n)\otimes
\openone_{2N}
\]
ensures convergence of the Gaussian integral. Finally, the `source
field' $\hat J$ is block diagonal in flavour space,
\[
\hat
J = {\rm bdiag\;}(\hat J_1,\dots,\hat J_n)\otimes
\sigma_3^{\rm bf},
\] 
where $\hat J_m = \{(\hat J_m)_{ij} \}, m=1,\dots,n$ are arbitrary
matrices on the lattice, and $\sigma_3^{\rm bf}$ is the Pauli matrix in
the space of grading or $\sigma$-indices.

The significance of the functional integral (\ref{Z1}) lies in the
fact that averaged products of single-particle Green functions can be
obtained from it by differentiation with respect to matrix elements of the 
source field $\hat J$: As with the application of supersymmetric
field integrals to the analysis of Green functions of conventional
disordered systems, we have the identity
\begin{equation}
  \frac{\delta^n Z[\hat J]}{\delta (J_1)_{i_1j_1}\dots \delta
    (J_n)_{i_nj_n}} = 2^n  \left\langle\prod_{m=1}^n
    G_{j_m,i_m}(z_m)\right\rangle.
\end{equation}
In the following it is important to notice the existence of a
symmetry of the functional (\ref{Z1}) which is directly inherited from
the chirality of the Hamiltonian: Consider a linear transformation of
the integration variables,
\begin{eqnarray}
  \label{chiral_psi}
&&\psi_i \to T\psi_i,\qquad \bar \psi_i \to \bar \psi_i
\bar T,\qquad i\in A,\nonumber\\
&&\psi_j \to V\psi_j,\qquad \bar \psi_j \to \bar \psi_j
\bar V,\qquad j\in B,
\end{eqnarray}
where $T,\bar T,V,\bar V \in {\rm
  Gl}(n|n)$ are arbitrary invertible supermatrices. In the limit $\hat
z \to 0$, the (source free, $\hat J=0$) action remains invariant
provided that the transformation matrices  are related to each other by
\begin{eqnarray}
\label{TV_connect}
 \bar T s V = s,\qquad
  \bar V s T = s.
\end{eqnarray}
In other words, the $\hat z=0$ functional integral possesses a global ${\rm
  Gl}(n|n) \times {\rm GL}(n|n)$  symmetry. Notice that for a
general non-chiral Hamiltonian (i.e. a Hamiltonian including diagonal
blocks in the representation (\ref{chiral1})), the symmetry of the
  functional integral 
collapses to ${\rm GL}(n|n)$. 

What is missing in the above argument is that the matrices $T,\bar
T,V$ and $\bar V$ can in fact not be chosen independently: The
relation $\bar \psi_{\rm B}= \psi_{\rm B}^\dagger$ implies that $\bar
V_{\rm BB} = V_{\rm BB}^\dagger$ and $\bar
T_{\rm BB} = T_{\rm BB}^\dagger$. Consequently, in the
${\rm BB}$-sector of the theory, the global symmetry collapses from
${\rm GL}(n)\times {\rm GL}(n)$ to just ${\rm GL}(n)$. Later on we will
see that the symmetry manifold is even smaller than that. We will come
back to discussing the consequences of such reduction mechanisms after
the theory has been subjected to the colour-flavour transformation.

Having identified the basic symmetries of the functional integral, the
next conceptual step in the construction of the field theory is the
average over the bond matrices $U_{ij}$. At this stage the
`colour-flavour transformation', mentioned above, is applied. It
effectively trades the group integrals over the $U_{ij}$ for
integrations over a set of auxiliary field variables $Z_{ij}$:
Applying the identity Eq.~(3) in Ref.~\cite{Zirnbauerb}, i.e. the
transformation identity for a single ${\rm U}(N)$ integration, to all
links of the lattice we obtain
\begin{eqnarray}
\label{Z2}
&&Z[\hat J]= \int {\cal D}(Z,\tilde Z) {\cal
  D}(\psi,\overline\psi)
e^{N\sum_{\langle i\in A, j\in B \rangle}\rm str \, ln \,
(1-Z_{ij}\tilde Z_{ij})}\times\nonumber\\
&&\hspace{1.0cm}\times e^{i\sum\limits_{i\in A} 
\overline\psi_i s^{1/2}
 \left(\hat{z} +\sum_{j\in N_i} Z_{ij}\right) s^{1/2} \psi_i+
i\sum\limits_{j\in B} 
\overline\psi_j s^{1/2}
 \left(\hat{z} +\sum_{i\in N_j} \tilde Z_{ij}\right) s^{1/2}
  \psi_j+
i\sum_{k,l}\overline\psi_k s^{1/2}
 \hat J_{kl} s^{1/2}
  \psi_l }.
\end{eqnarray}
Here 
\begin{equation}
\label{Z_domain_global}
  Z_{ij}= \{(Z_{ij})_{aa'} \}\in {\rm GL}(n|n),\qquad  \tilde
  Z_{ij}= \{(\tilde 
  Z_{ij})_{aa'} \} \in {\rm GL}(n|n) 
\end{equation} 
are $2n\times 2n$-dimensional supermatrices living on the
directed link from the $A$ site $i$ to the nearest neighbour $B$-site $j$. (By
`directed' we mean that there is no $Z_{ji}$.) The symbol $\sum_{j\in N_i}$
stands for a summation over all $j$'s that are connected with
$i$ and `str' is the standard supertrace~\cite{VWZ}. 

In (\ref{Z2}), the integration $\int {\cal D}(Z,\tilde Z)$ over all
pairs $(Z_{ij},\tilde Z_{ij})$ is subject to a number of
constraints: The  fermion-fermion blocks of the matrices
$Z_{ij}$ and $\tilde Z_{ij}$ are anti-Hermitian adjoint to each other,
\begin{equation}
\label{Z_domain_FF}
Z_{ij,\rm FF}^{\vphantom{\dagger}}= - \tilde Z_{ij,\rm FF}^\dagger.
\end{equation}
In the boson-boson block, a similar but more subtle constraint needs
to be imposed. In the original formulation of the colour-flavour
transformation~\cite{Zirnbauerb} the ${\rm BB}$-blocks of $Z_{ij}$ and
$\tilde Z_{ij}$ were chosen to be Hermitian adjoint, $Z_{ij,{\rm
    BB,orig.}} = \tilde Z_{ij,{\rm BB,orig.}}^\dagger$.  As will
become clear later on, this convention is not suitable for the
analysis of the RF-problem. We rather need to chose a different
integration manifold, where
\begin{equation}
  \label{Z_domain_BB}
  \tilde Z_{ij,{\rm BB}}\equiv
Z_{ij,{\rm BB}}^\ddag\not=Z_{ij,{\rm BB}}^\dagger.
\end{equation}
In order to define the meaning of the symbol '$\ddag$', we first note
that $Z_{\rm BB} \in {\rm GL}(n)$. Furthermore, due to ${\rm GL}(n) = {\rm
  Gl}(n)/{\rm U}(n) \times {\rm U}(n)$, any matrix $Z_{\rm BB}$
can be decomposed as 
\begin{equation}
\label{Z_BB_factor}
Z_{\rm BB} = e^{iH'} e^{H},
\end{equation}
where both $H$ and
$H'$ are Hermitian (implying that the factor $e^{H}\in {\rm
  U}(n)\backslash {\rm GL}(n)\simeq {\rm GL}(n)/{\rm
  U}(n)$ whilst $ e^{iH'} \in {\rm U}(n)$). The $\ddag$-operation is then 
defined by
\begin{equation}
  \label{ddag_def}
Z^\ddag = ( e^{iH'} e^H)^\ddag \equiv e^{-H} e^{iH'},
\end{equation}
i.e. it changes the signs of the exponents in a manner that is opposite from
the standard Hermitian adjunction `$\dagger$'.  As detailed in
Appendix~\ref{sec:boson_int}, the new integration domain can be reached
from the original one by analytic continuation.  Hence, switching
between the two conventions does not affect the result of the
integral. Finally, the Grassmann valued components, $Z_{ij,{\rm BF}},
Z_{ij,{\rm FB}}, \tilde Z_{ij,{\rm BF}}, \tilde Z_{ij,{\rm FB}}$ are
independent $n\times n$-matrices.

\begin{figure}
\begin{center} 
\epsfig{file=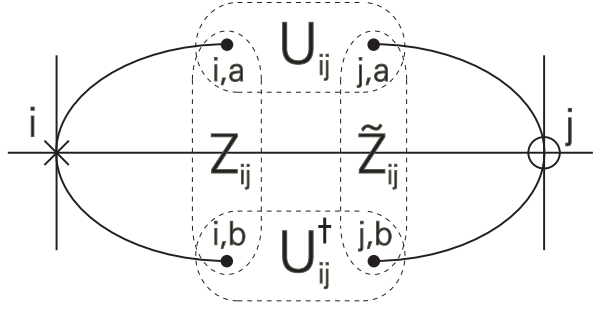,height=4cm}
\end{center}
\caption{\label{fig:cft} Visualisation of the different coupling of
  the $U$'s and the $Z$'s, respectively, to the field variables $\psi$
  (represented   by the solid dots).}
\end{figure}

Conceptually, the transformation (\ref{Z2}) has a
status similar to the Hubbard-Stratonovich transformation employed in
the standard construction of field theories for systems with Hermitian
disorder~\cite{Efetovsbook}: The motivation for introducing the $Z$-fields 
is that they
couple to the fields $\psi$ in a way different from the coupling $\bar
\psi U \psi$ between the $\psi$'s and the unitary disorder (see fig.
\ref{fig:cft}). Qualitatively, the $Z$'s connect pairs of
fields that, by construction, represent segments of `paired paths'.
As discussed in section \ref{sec:qualitative}, segments of this
structure are stabilised by mechanisms of quantum interference,
implying that the $Z$'s connect to a sector of the theory which
contains information on long range behaviour. In addition, the $Z$'s
couple non-trivially to the `flavour' space thereby potentially
allowing for the description of quantum interference between different
flavour species ($\leadsto$ different Green functions).  The
subsequent analysis of the field theory will essentially amount to
extracting the low energy interference stabilised sector contained in
the $Z$-functional.

The first step towards an effective long range model is to integrate
out the $\psi$-fields. At this stage one runs into an unexpected
problem, viz. the Gaussian integration over the bosonic component of
the $\psi$'s does not converge for general complex energy
arguments $z_m$. In other words, it looks like the orders of
integrations over $Z$ and $\psi$ cannot be exchanged which would imply
that the colour-flavour transformation is useless.

The problem can be surmounted by analytically continuing the arguments
$z_m$ from their natural definition range $z_m \equiv \epsilon_m \pm
i\delta$, $\delta$ positive infinitesimal, $\epsilon_m$ a real
`energy' argument, deeply into the complex plane: $\epsilon_m \pm
i\delta \to \epsilon \pm 2id$.  As shown in Appendix
\ref{sec:convergence}, this manipulation cures the problems associated
with the $\psi$-integration.  Moreover, we will see shortly that the
remaining integration over $Z$ is not plagued by convergence problems
(as long as the $z_m$'s stay away from the real axis.) As a
consequence the shift into the complex plane can be redone
immediately after the $\psi$'s have been integrated out.

Integrating the partition function (\ref{Z2}) over the field $\psi$
(the manipulation of the energy arguments mentioned above being understood),
we obtain the functional integral 
\begin{eqnarray}
\label{Z3}
&&Z[\hat J]=\int {\cal D}(Z,\tilde Z)e^{-S[Z,\tilde Z]}X[\hat J],\nonumber\\
&&\qquad S[Z,\tilde Z]= 
-N\sum_{\langle i\in A, j\in B
    \rangle}{\rm str \, ln \,} 
(1-Z_{ij}\tilde Z_{ij})+\nonumber\\
&&\hspace{2.5cm}+N\sum\limits_{i\in A} {\rm str \, ln \,}
\left(\hat{z} +\sum_{j\in N_i} Z_{ij}\right)+N\sum\limits_{j\in B} {\rm
  str \, ln \,} 
 \left(\hat{z} +\sum_{i\in N_j} \tilde Z_{ij}\right),
\end{eqnarray}
where $X[\hat J]$ represents certain pre-exponential contributions
resulting from an expansion in powers of the source field $\hat J$ to
$n$th order.  Since, in this study, the focus will be on an analysis of
the general structure of the theory rather than on the computation of
specific correlation functions, we will henceforth ignore
this contribution. (To actually compute correlation functions one may
proceed in a standard manner, i.e. keep track of $X[\hat J]$,
differentiate with respect to $\hat J$ at any convenient stage, and eventually
carry out the $Z$-integral weighted with the effective action derived
below.)

\section{Effective Field Theory}
\label{sec:effective}
So far all manipulations have been exact. In order to make further
progress towards an effective long range theory, we next subject the
functional (\ref{Z3}) to a standard saddle point analysis, that is we
seek for solutions of the equations
\begin{eqnarray*}
\left.\frac{\delta S[Z,\tilde Z]}{\delta Z_{ij}}\right|_{Z=Z_0,\tilde Z=\tilde Z_0}&=&0,\\
\left.\frac{\delta S[Z,\tilde Z]}{ \delta \tilde
    Z_{i'j'}}\right|_{Z=Z_0,\tilde Z=\tilde Z_0}&=&0.
\end{eqnarray*}
Anticipating that spatial fluctuations of the fields $Z$ and $\tilde
Z$ will be energetically penalised, we make an ansatz where the saddle
point fields $Z_0$ and $\tilde Z_0$ connected to the $A$ and the $B$
sublattice, respectively, are spatially constant. Furthermore, the
fact that the action is diagonal  in the composite index
$a=(m,\sigma)$, permits one to seek for solutions of the form
$(Z_0)_{aa'},(\tilde Z_0)_{aa'}\propto \delta_{aa'}$. 

Substituting this Ansatz into the functional integral and
varying the action we obtain the equations
\begin{equation}
\label{spe}
\tilde Z_0 = Z_0 = -\frac{x}{2} \hat z \pm x \left(-1 +
  \frac{x^2}{4}\hat z^{2}\right)^{1/2},
\end{equation}
where $x \equiv (2d -1 )^{-1/2}$.  Close to the middle of the band,
$z_m\ll1$, the solutions of (\ref{spe}) compatible with the
analyticity requirements imposed by the logarithms in the action are
given by
\begin{equation}
\label{diag_sp}
\tilde Z_0 = Z_0 \simeq ix  s.
\end{equation}

As in the construction of the field theory of the weakly disordered
sublattice model~\cite{gade}, the diagonal matrices (\ref{diag_sp})
are not the only solutions of the saddle point equations. The isotropy
of the $\hat z=0$ action in the space of $a$-indices rather implies
that the set of solutions is highly degenerate. In order to identify
the structure of the entire saddle point manifold we proceed in a
standard manner and investigate how a general chiral symmetry
transformation (\ref{chiral_psi}) affects the diagonal saddle point
solutions. From the structure of the action (\ref{Z2}) it follows that
under (\ref{chiral_psi})
\begin{eqnarray}
\label{nondiag_sp} 
  Z_0 = ix s^{-1/2}s^{-1/2} &\to&ix s^{-1/2} \bar T T
  s^{-1/2}, \nonumber\\
  \tilde Z_0 = ix s^{-1/2}s^{-1/2} &\to&ix s^{-1/2} \bar V V s^{-1/2}
  \stackrel{(\protect \ref{TV_connect})}{=}ix s^{1/2} (\bar T T)^{-1}
  s^{1/2}.
\end{eqnarray}
Since the action of the theory is invariant under global chiral
symmetry transformations, all pairs $(Z_0,\tilde Z_0)$ of the
structure (\ref{nondiag_sp}) are solutions of the saddle point
equations, too. Notice that in (\ref{nondiag_sp}) the ${\rm
  Gl}(n|n)$-matrices $T$ and $\bar T$ appear in a product combination.
Taking into account that ${\rm GL}(n|n)\cdot {\rm GL}(n|n)={\rm
  Gl}(n|n)$, we conclude that the full extent of the saddle point
manifold is given by ${\rm GL}(n|n)$.\footnote{This is an example of what 
  in QCD is known as the
  phenomenon of `chiral symmetry breaking': Neglecting quark masses,
  the microscopic Yang-Mills action (corresponding to our $\psi$-action
  in the small $z$-limit), is invariant under ${\rm G} \times {\rm G}$
  where $G$ is a symmetry group whose detailed structure depends on
  the flavour content of the theory etc. (In our formalism, its role
  is played by ${\rm GL}(n|n)$.) Now, in the {\it effective} low
  energy QCD Lagrangian, the microscopic chiral symmetry is
  spontaneously broken down to ${\rm G} \times {\rm G}\to {\rm G}$,
  very much like the symmetry group of our low energy $Z$-functional
  is ${\rm GL}(n|n)$ rather than ${\rm GL}(n|n)\times {\rm GL}(n|n)$.}

Eq.~(\ref{nondiag_sp}) defines the maximum manifold of solutions of
the saddle point equations. In order to complete the determination of
the global structure of the theory, we next need to explore how the
manifold (\ref{nondiag_sp}) intersects with the domain of the field
integration (as specified by (\ref{Z_domain_global}),
(\ref{Z_domain_FF}) and (\ref{Z_domain_BB})). To this end, we
represent a general pair of matrices $Z$ and $\tilde Z$ according to
\begin{equation}
\label{Z_vs_PT}
  (Z,\tilde Z)\equiv ix(PT,T^{-1}P), 
\end{equation}
where $P,T\in{\rm GL}(n|n)$ and we have omitted the site index $(ij)$
for notational transparency.

In order to give the reparameterisation (\ref{Z_vs_PT}) some meaning,
we a) have to be specific about the internal structure of the factor
matrices $T$ and $P$, and b) tell how the integration measure ${\cal
  D}(Z,\tilde Z)$ is affected by the transformation. As for a), the
restriction (\ref{Z_domain_FF}) leads to the relation
\begin{equation}
  \label{PT_FF}
  P_{\rm FF}\in {\rm GL}(n)/{\rm U}(n),\qquad T_{\rm FF}\in {\rm U}(n).
\end{equation}
Similarly, the decompositions (\ref{Z_domain_BB}) and
(\ref{Z_BB_factor}) imply that
\begin{equation}
  \label{PT_BB}
  P_{\rm BB}\in {\rm U}(n),\qquad T_{\rm BB}\in {\rm GL}(n)/{\rm U}(n).
\end{equation}
The off-diagonal blocks $P_{\rm BF},P_{\rm FB},T_{\rm BF}$ and $T_{\rm
  FB}$ are independent Grassmann variables.  Finally, counting the
number of independent parameters, we see that the $(P,T)$ manifold
covers the entire $(Z,\tilde Z)$ domain.

As for the question of the measure, first note that Eqs.
(\ref{nondiag_sp}) and (\ref{Z_vs_PT})  imply that the saddle point
configurations of the theory are specified by $P= \openone$.
Consequently, excitations in $P$ are massive, and can be treated in a
quadratic approximation. Due to the fact that close to $P=\openone$ the
integration measure is approximately flat, we are spared addressing
point b) explicitly. More precisely, the measure of the
$P$-integration is flat Euclidean, whilst the $T$-integration is over
the invariant measures on the manifolds (\ref{PT_FF}) and
(\ref{PT_BB}), respectively.

Eqs. (\ref{Z_vs_PT}), (\ref{PT_FF}) and (\ref{PT_BB}) contain the
answer to the question of the general structure of the target manifold
of the field theory raised above: The maximal manifold of solutions of
the saddle point equation is given by ${\rm GL}(n|n)$.  In the ${\rm
  BB}$-sector, the intersection of this manifold with the actual
domain of the field integration is given by ${\rm GL}(n)/{\rm
  U}(n)\subset {\rm GL}(n)$, and in the ${\rm FF}$-sector, by ${\rm
  U}(n)\subset {\rm GL}(n)$. The Grassmann sectors are not restricted.
We note that these global structures have been worked out
earlier~\cite{Zirnbauera}, albeit in a more general manner based on
Cartan's complete classification of symmetric spaces. Apart from the
fact that it is not obvious how that more abstract line of reasoning
can operationally be adapted to the RF-model, the present treatment
has the advantage of encompassing the massive sector of the theory
(the $P$'s). As we will see later on, the RF-problem differs from
standard (non-chiral) field theories of disordered systems, in that
the massive $P$'s have a significant impact on the final structure of
the functional integral.

Finally, we note that the original boson-replica~\cite{gade} and
fermion-replica~\cite{tsvelik1} based
approaches chiral disordered model systems led to field
theories over ${\rm GL}(Rn)/{\rm U}(Rn)$ and ${\rm U}(Rn)$, respectively. 
Here $R\to 0$ denotes the
number of replicas.  This field manifold can be interpreted as the
replica analogue of the ${\rm BB}$-sector of our supersymmetric
construction.

We next employ the saddle point decomposition (\ref{PT_FF}) and
(\ref{PT_BB}) to reduce the exact functional (\ref{Z3}) to a simpler
effective functional describing the long range behaviour of the model.
Technically, the procedure will be based on a continuum approximation
in combination with a gradient expansion around the spatially constant
saddle point manifold. It is important to realize that this evaluation
scheme is, in fact, problematic: As compared to analogous treatments
of {\it weakly} disordered models, the gradient expansion is not
stabilised by the presence of the small parameter `disorder
strength/Fermi energy'. Apart from the external parameters $z_m$, all
energy scales are rather of the same order: ${\cal O}(1)$. With the
non-Abelian versions of the RF-model, $N>1$, the situation is less
problematic, the reason being that the mean field analysis is
stabilised by the parameter $1/N$.  However, for the $N=1$ random {\it
  magnetic} flux model, the expansion around the mean field
configurations is not controlled by a small parameter. In this
situation all we can do is provide some independent physical evidence
for the belief that the field theory constructed below remains valid
all the way down to $N=1$.\footnote{We remark that this extrapolation has
the same status as the applicability of the Pruisken action to the Integer
Quantum Hall transition. Strictly, the field theory can be formally justified
only in the limit of high Landau level index, or large conductance. Yet, 
studies~\cite{Zirnbauerc} of the phenomenological Chalker-Coddington model, 
corresponding to a single Landau band, indicate that the same action controls 
the fixed point of the Hall transition when the conductance is $O(e^2/h)$.} 
Some arguments to that effect will be given towards the end of the paper.

Keeping these words of caution in mind, we now turn to the discussion of the
continuum expansion of the theory.  On general grounds, it may be
anticipated that after integration over massive modes, the effective
action takes the form
\begin{equation}
  \label{S_eff_schematic}
  S[P,T] \stackrel{\langle \dots \rangle_P}{\longrightarrow}S_{\rm eff}[T]
  = S_z[T]+S_{\rm fl}[T] + S_{\rm m}[T],
\end{equation}
where $S_z[T]$ is the contribution due to finite energies $\hat z$,
$S_{\rm fl}[T]$ the action associated with fluctuations of the Goldstone 
modes (the $T$'s), and $S_{\rm m}[T]$ is a residual action induced by the
interaction between massive and Goldstone modes, respectively. We next
assume that the field configurations relevant to the long range
physics are smooth (differentiable). The different contributions to
the action can then be computed by a gradient expansion followed by a 
continuum limit. Albeit conceptually straightforward, the explicit
formulation of this program happens to be somewhat cumbersome. For
this reason, the derivation of the different contributions to the
action has been deferred to Appendix \ref{sec:gradient}.  As a result
we obtain 
\begin{eqnarray}
  \label{Seff}
&&  S_{\rm eff}[T] = -\frac{1}{2}
\int \Bigg( \frac{a^{2-d}}{b} {\,\rm str}( \partial
    T^{-1} \partial T ) + i\frac{2N(2d-1)^{1/2}a^{-d}}{4d}  {\rm
      str\,} (\hat z (T+T^{-1})) + \nonumber\\
&&\hspace{4.0cm} +\frac{a^{2-d}}{c} {\,\rm
    str\,}( T^{-1}\partial T) {\,\rm str \,}( T^{-1}\partial
  T)\Bigg)+S_{\rm b}[T],
\end{eqnarray}
where $a$ represents the lattice spacing, $b^{-1} = N/4d$,
$c^{-1}=C/8d$, and $C$ denotes a numerical constant defined in
Eq.~(\ref{Cdef}).  Further, the discrete lattice summations $\sum_i$
have been converted to integrals over smooth field configurations $T$.
Eq.~(\ref{Seff}) represents the main result of this paper. $S_{\rm
  b}[T]$ represents a boundary action whose structure depends
on the geometry of the lattice. For a $d$-dimensional hypercubic
lattice with $N_i$ sites in direction-$i$,
\begin{equation}
  \label{Sb}
  S_{\rm b}[T]=\frac{N}{2^d}\sum_{s_1,\dots,s_d=0,1} \prod_{i=1}^d
(-)^{(N_i+1)s_i} {\rm str \, ln\,} (T(s_1 L_1,\dots,s_dL_d)).
\end{equation}
Notice that the boundary action depends on the `parity' of the
site-numbers, i.e. on whether or not the number of sites in the
different extensions of the lattice is even on odd. Both the
implications of this fact and the topological content of the boundary
action will be discussed below. 

\section{Analysis of the Field Theory}
\label{sec:eval}
What kind of information can be extracted from the effective action
(\ref{Seff})? A comparatively complete answer to that question may be
given in the `zero dimensional' case, i.e. the case of systems of linear
extension $L$ in a regime where all energy parameters are such that
$z_m \ll a^2/L^2$. (Recall that all energies are measured in units of
the modulus of the hopping matrix elements.)  The denotation
`zero dimensional' is motivated by the fact that, as in the case of
conventional field theories of disordered systems, the boundary
conditions to be imposed on $\partial T$ imply that the kinetic energy
associated with spatial fluctuations is quantised: For constant field
configurations it vanishes, while the first non-constant configuration has a
fluctuation energy of ${\cal O}(N a^2/L^2)$, the `Thouless energy' of
the system\footnote{ That $a^{2}/L^2$ plays the role of the Thouless energy
  $E_c\sim v_F l/L^2 = v_F^2 \tau /L^2$ ($v_F$: Fermi velocity, $l$:
  elastic mean free path, $\tau$: elastic scattering time) of weakly
  disordered systems can be seen from the following rough estimate:
  Remembering that all energies are measured in units of the hopping
  matrix element, $v_F \sim a$ in the middle of the band.  Further,
  for strong disorder, $\tau \sim 1$ is of the order of the band
  width. This gives $E_c \sim a^2/L^2$.
}. (Notice that in $d=2$, the Thouless energy is
small of ${\cal O}(\Delta)$, where $\Delta$ is the mean level spacing.
Thus, unlike in weakly disordered $2d$ systems, there is no extended
region separating the onset of the zero mode regime $z\sim E_c\equiv
a^2/L^2$ from the fundamentally quantum regime $z\sim \Delta$.) For
energies $z_m \ll a^2/L^2$ spatially fluctuating configurations are
effectively frozen out and the action simplifies to
\begin{equation}
  \label{S0mode}
  S_{0}[T] = -i {\pi\over 2}\rho_0 
   {\rm      str\,} (\hat z (T+T^{-1}))+S_{\rm b}[T],
\end{equation}
where $T \in {\rm GL(n|n)}$ is a single matrix (the zero mode), and 
\begin{equation}
  \label{rho_0}
  \rho_0 = \frac{N(2d-1)^{1/2}L^2}{2\pi da^2}.
\end{equation} 

To illustrate how the zero mode action can be applied to the analysis
of physical observables, we consider the DoS (\ref{DoS_def}) in the
zero dimensional limit. Applying the general recipe for the computation of 
Green functions outlined above to the source field $\hat J=\sigma_3^{\rm bf}$, 
the DoS is represented by 
\begin{equation}
  \label{DoS1}
  \rho (\epsilon)  = \frac{\rho_0}{4}{\, \rm
  Re\,}\int dT e^{-S_0[T]} {\,\rm 
  str\,}((T+T^{-1})\sigma_3^{\rm bf}).
\end{equation}
The integral depends sensitively on the presence of the boundary term 
$S[T]_{\rm b}$ which in turn depends on the parity of the lattice. If the 
number of sites is odd (i.e. the number of lattice sites along {\em each and 
every} direction is odd), the boundary term takes the form
\begin{equation}
S_{\rm b}[T]={\rm str}\ln T,
\end{equation}
otherwise the term vanishes. In both cases the explicit computation of the
zero-mode integral (\ref{DoS1}) for the particular case of $n=1$ is detailed 
in Appendix~\ref{sec:0-mode}. Explicit calculation shows that, for a lattice 
with an even number of sites, the DoS takes the universal form
\begin{equation}
  \label{rho}
  \rho(\epsilon) =\frac{\pi^2 \epsilon \rho_0^2}{2} \left({\rm
  J}_0^2(\pi \epsilon \rho_0) + {\rm J}_1^2(\pi \epsilon \rho_0)\right),
\end{equation}
while if the number of sites is odd, 
\begin{equation}
\label{rho2}
\rho(\epsilon)=\frac{\pi^2 \epsilon \rho_0^2}{2} \left({\rm J}_N^2(\pi\epsilon 
\rho_0)-{\rm J}_{N-1}(\pi \epsilon \rho_0){\rm J}_{N+1}(\pi \epsilon \rho_0)
\right),
\end{equation}
where ${\rm J}_N$ denotes the Bessel function of order $N$. 

In fact, both of these results have been anticipated in the literature.
Firstly, Eqs.~(\ref{rho}) and (\ref{rho2}) have been derived earlier within 
the framework of a supersymmetric non-linear $\sigma$-model for 
the chiral random matrix ensemble ChGUE~\cite{ast,v4}, and within 
pure random matrix theory~\cite{v1,v2,v3,ns}. Indeed, both expressions
have been discussed in their connection with chiral symmetry breaking of the 
QCD Dirac operator~\cite{v1,v2,v3,v4}. Moreover, in the present context, 
the validity of Eq.~(\ref{rho}) has been investigated 
numerically~\cite{Avishai2}. 

In fact, the coincidence between the low 
energy RF-field theory and the non-linear $\sigma$-model for chiral random 
matrices is not restricted to the mean DoS but rather extends to arbitrary 
$n$-point correlation functions. Equivalently, it can be said that in the 
low energy regime the RF-model is described by chiral random matrix theory, a 
fact that could have been anticipated from symmetry arguments alone. 

Eq.~(\ref{rho}) also contains the answer to the question
of what happens to the DoS upon leaving the domain of perturbation
theory (c.f. Eq.~(\ref{DoS_path}) and below): Ultimately (i.e. on a
scale comparable with the mean level spacing) it drops down to zero.
Notice, however, that the zero dimensional analysis cannot exclude the
existence of divergent behaviour prevailing down to energy scales
$\sim E_c$ (see below).

At the same time, Eq.~(\ref{rho}) suggests that on energy scales $\epsilon >
\rho_0^{-1}$, the DoS approaches the constant value $\rho_0$. This
prediction must be taken with some care: In $d=2$ it is meaningless
because the domain of applicability of the zero mode approximation is
set by $\rho_0^{-1}$ itself and the present analysis has nothing to
say about energies larger than that. For $d>3$, $\rho_0$ should have
the significance of a `mean' DoS of the random flux model. Although,
indeed, Eq. (\ref{rho_0}) agrees parametrically with approximate
expressions for the band center DoS of large dimensional
hypercubic lattices\cite{MuHa2}, it is not clear whether $\rho_0$ has
any true physical significance for the physics of the RF models. The
reason is that (\ref{rho_0}) represents a bare parameter which can be, in
principle, subject to substantial renormalization. As will be discussed below,
the extent to which the parameter changes under the RG flow is, at
present, not clear.

Leaving the zero-mode regime and returning to the more complicated,
truly extended case, we first note that the field theory described by
(\ref{Seff}) is not entirely original but rather has a number of
closely related precursors:
\begin{itemize}
\item Gade's analysis of the weakly disordered sublattice model led to
  a boson replica version of the present model, i.e. a theory over
  fields $T\in {\rm GL}(nR)/{\rm U}(nR)$, where $R\to 0$ is the number
  of replicas. The action for these fields coincided with
  (\ref{Seff}), save for the important difference that, due to the
  weakness of the disorder, the coupling constant $b^{-1}$ was
  parametrically larger than one.
\item In high energy physics the action (\ref{Seff}) has been suggested on
  phenomenological and symmetry grounds as relevant for the effective
  description of partially quenched QCD~\cite{bernard-golterman}. In that
  context, the base manifold is $4+1$-dimensional whilst the fields
  $T\in U(n_f+1|1)$, where $n_f$ is the number of quark flavours. 
\item A similar field theory (albeit without a 
  $({\rm tr}(T^{-1}\partial T))^2$ operator) was derived in connection
  with a model of Dirac fermions subject to a random non-Abelian
  vector potential~\cite{tsvelik2}. That the random Dirac model is closely
  related to the RF-problem considered here follows from the fact
  that, when put on a lattice, Dirac fermions are equivalent to free
  fermions subject to an average $\pi$-flux per lattice cell. In other
  words, the Dirac model corresponds to a weak random flux model
  centered around an average $\pi$-flux background.

  As for the field theoretical description, apart from being defined on
  a different field manifold (${\rm SU}^c(N)/{\rm SU}(N)$, where ${\rm
    SU}^c(N)$ is the complex extension of ${\rm
  SU}(N)$\footnote{Notice that for $T\in {\rm SU}^c(N)/{\rm SU}(N)$,
  ${\;\rm det\,}(T)=1 \Rightarrow {\rm tr\,}(T^{-1}\partial T)=0$ which
  explains the absence of the  $({\rm tr}(T^{-1}\partial T))^2$
  operator in this field theory.}), the model differs from (\ref{Seff}) in 
  that it contains a Wess-Zumino-Novikov-Witten (WZNW) operator (however, see
  Ref.~\cite{foot_dwave}). While the symmetries of the 
  fields $\in {\rm GL}(n|n)$ would, in principle, admit a supersymmetric
  version of a WZWN operator, we believe that the
  presence of a WZNW action is induced by the Dirac nature of the clean
  spectrum implying that there is no room for such a contribution 
  in our model.
  (C.f. the absence of a Pruisken term in the action
  of a weak disordered electron gas subject to a random, {\em zero average} 
  magnetic field.) To check that conjecture explicitly, we have computed
  the action of an ${\rm SU}(2)$ instanton type configuration defined
  in the FF-sector of the theory.  No imaginary contribution to the
  action was obtained, a signature for the absence of a WZWN
  contribution. For the application of a similar recipe to determine
  the coupling constant of topological contributions to the action,
  see Ref.~\cite{Zirnbauerc}. 
\item Finally, a fermion-replica analogue of the action (\ref{Seff}) was 
  obtained 
  in connection with a lattice model of a disordered $d$-wave superconductor
  \cite{tsvelik1,Senthil98,Altland_dwave}. The connection to the RF-model 
  follows from the fact that the clean limit of the $d$-wave model is 
  governed by the
  presence of Dirac type excitation spectra and can therefore be
  effectively described in terms of a model qualitatively analogous to 
  that of $\pi$-flux lattice fermions.

\end{itemize}

Various conclusions regarding the physical behaviour of the RF-model,
most notably about its localization behaviour, can be directly
inferred from Ref.~\cite{gade}. There it was shown that the conductance 
of the weakly disordered $2d$ model at the band center (which is
essentially determined by the coupling constant $b^{-1}$) did not
change under one-loop perturbative renormalization. This observation
suggests that a non-localised state might exist in the middle of the band.
Since the stability of the perturbative RG merely relies on the
smallness of the parameters $b,b^2/c \ll 1$, its results can be
straightforwardly carried over to the $N \gg 1$ non-Abelian RF
model: The one-loop renormalization indicates that for $N \gg 1$ the
strongly disordered RF model exhibits metallic behaviour at the band center.

It is natural to ask as to what extent this prediction survives the
limit $N\to 1$. Actually, one may adopt an even more hesitant attitude
and ask whether the basic {\it structure} of the action (\ref{Seff}) is
stable as $N\to 1$ (c.f. comments made in Appendix
\ref{sec:gradient} in connection with the gradient expansion). We
believe that the answer to the second question is positive whilst the
answer to the first one remains, to some extent, a matter of faith.

Addressing the second issue first, we note that, by a standard
argument of operator relevancy, the infrared continuum action should
be dominated by operators with the least number of derivatives. Of
these, there are only two invariant combinations that are compatible
with the basic symmetries of the model, namely the two operators
appearing in (\ref{Seff}). (E.g.  the presence of an operator like
$\sim {\,\rm str}( \partial_i T^{-1} \partial_j T ), i\not=j$ is ruled
out by virtue of reflection symmetry, etc.). Thus, the basic structure
of the action is dictated by symmetry alone. As for the coupling {\it
constants} multiplying the operators of the action, the absence of
dimensionful parameters in the RF problem fixes their value up to
numerical factors about which we need not be concerned.

Now, turning to the more serious question of the validity the RG
results in the Abelian case, from the results presented here, there appears 
to be only one useful test of consistency: The existence of a metallic state 
for $N=1$ with strong disorder would at least be compatible with the 
behaviour in the limiting cases: $N=1$ and weak disorder, and $N\gg 1$ with 
arbitrary disorder. Moreover, this conclusion does also seem to be consistent 
with the most recent numerical investigations~\cite{Furasaki}. More concrete 
analytical evidence in favour of band center delocalisation does not seem 
available. 

Besides the conductance, the DoS of the extended model also behaves
in an unusual manner: The analysis of Ref.~\cite{gade} predicts divergent
behaviour upon approaching the band center, where the detailed
functional form of the divergence depends on the dimensionality of the
system. In particular, in two dimensions the RG analysis predicts
\[
\rho(\epsilon) \sim \frac{e^{-\kappa
    \sqrt{-\ln(|\epsilon|)}}}{|\epsilon|}.
\]
The scale below which the $|\epsilon|^{-1}$ divergence begins to
dominate over the exponential factor is determined by the constant
$\kappa$. Unfortunately the dependence of that constant on the
relevant parameters of the model can
not be reliably extracted from the perturbative RG analysis, i.e. the
only statement that can safely be made is that a divergence forms {\it
  somewhere} upon approaching the band center. Ultimately, the
divergent behaviour is cut-off by the onset of the
non-perturbative ergodic regime discussed above. A quantitative
understanding of both the formation of the divergence and its truncation
in the close  vicinity of the band center would necessitate a
combination of an RG analysis with a non-perturbative treatment of the
zero mode contribution, a program that is well beyond the scope of the
present paper.

A third piece of information that can be extracted from Ref.~\cite{gade} is
how finite energy arguments $z_m\equiv \epsilon_m \pm i0$ affect the
behaviour of the model: In the spirit of the RG, finite energies have
the significance of  relevant perturbations. Thus, the band center
represents a critical manifold whilst, for finite $z_m$, the model
flows into a different universality class. More precisely, the band
center field theory --- a model of universality class AIII in the
classification of Ref.~\cite{Zirnbauera} or, more commonly, a model of 
ChGUE-symmetry --- develops into an ordinary GUE type model. This is
accompanied by a reduction of the symmetry manifold from ${\rm
  Gl}(n|n)$ to the standard GUE manifold ${\rm GL}(n|n) / {\rm
  Gl}(n_+|n_+) \times {\rm GL}(n_-|n_-) $, where $n_++n_-=n$ and
$n_\pm$ is the number of energy arguments with positive/negative
imaginary part. (Note that this conclusion is in accord with the analysis 
in Ref.~\cite{Aronov} where a continuum random field model far away from the 
band center was shown to be described by a supersymmetric unitary model.)

In that context, one may ask how the degrees of freedom of the model
of reduced, unitary symmetry, commonly denoted by $Q$~\cite{Efetovsbook}, 
emerge from the fields $T$ of the present model upon crossing over to the
unitary regime. In fact the full RF model
could have been formulated in terms of $Q$-matrices (albeit $Q$'s of a
symmetry different from the standard unitary symmetry) from the
outset: I.e. by defining $Q=\tilde T \Lambda  \tilde T^{-1}$, where $\Lambda =
\sigma_3 \otimes s$, $\tilde
T=\exp (iW \sigma_2)$ and the $\sigma_i$ operate in sublattice space.
The matrices $\tilde T$ are then related to our $T$'s by the identity $T=\exp
(iW)$. The price to pay for this more common $Q$-formulation of the
theory would be a certain amount of redundancy: A description in terms
of structures $Q\sim \tilde T \Lambda T^{-1}$ is the natural one for
problems with {\it coset space} symmetry (e.g. the unitary coset space
${\rm GL}(n|n) / {\rm GL}(n_+|n_+) \times {\rm GL}(n_-|n_-) $).
However, the present problem has {\it group} type symmetry (${\rm
  Gl}(n|n)$) implying that a modelling directly in terms of group
degrees of freedom is more natural.
\section{Summary and Discussion}
\label{sec:summary}
Summarising, we have mapped both the Abelian ($N=1$) and non-Abelian
($N>1$) random flux model onto an effective continuum field theory.
The model differs in only two respects from the field theory obtained
earlier~\cite{gade} for band center lattice fermions subject to weak
off-diagonal disorder: It is a) a based on the formalism of
supersymmetry (as opposed to boson replicas), and b) governed by a
different coupling constant $b^{-1} \sim N \times {\cal O}(1)$ (as
opposed to the large value obtained in the weakly disordered case).

The supersymmetric formulation of the theory allows one to rigorously
establish contact with the formalism of chiral random matrix theory
\cite{v1,v2,v3,ns,hikami1,hikami2}, and to demonstrate that the
ergodic sector of the model is described by a random matrix ensemble
of appropriate symmetry. Apart from this point, however, the fact that
the present model is formulated in terms of supersymmetry, whilst the
earlier analysis~\cite{gade} utilised a replica formulation is
arguably incidental. What is, perhaps, more relevant is that both a
weakly and a maximally disordered model with sublattice symmetry are
described in terms of the same effective field theory. The only
difference lies in the value of the coupling constant $b^{-1}$. That
this constant takes a large/small value in the case of weak/strong
disorder follows readily from its significance as the bare conductance
of the model. Given that the two limiting cases (very weak/very strong
disorder) lead to the same effective action (\ref{Seff}), it is
tempting to conjecture that the present field theory universally
describes the long range behaviour of disordered sublattice models,
where the amount of disordering is reflected in the value of the
coupling constant $b^{-1}$.

Unfortunately, the small value of the coupling constant $b^{-1}$ in
the strongly disordered case prevents us from directly transferring
the results of the perturbative RG analysis of Ref~\cite{gade} to the RF case.
E.g. we cannot  exclude the existence of some fixed point
separating the delocalised regime for weak disorder from a potentially
localised regime of the fully random flux model. Thus, the present
analysis is not able to reliably answer the long-standing question of
the behaviour of the strongly disordered Abelian RF-model in the band
center.  What {\it can} be said with some confidence is that the
non-Abelian versions of the RF-model do exhibit long range
correlations ($\leftrightarrow$ delocalisation), this being a
consequence of the fact that $b^{-1} \propto N$ implying that the RG
is stable. (A more detailed account of the application of the present
formalism to the physics of systems describable in terms of
non-Abelian RF-models (or relatives thereof) will be presented
elsewhere.)

Finally, we mention that the present analysis of the RF-model can be
generalized so as to encompass other  lattice
problems with unitarily implemented disorder: Firstly,  models away from the 
`strong coupling limit' 
(models with a finite plaquette action like in (\ref{Slatt})) can be
dealt with at the expense of introducing extra `ghost components' in the
$\psi$-field\cite{Zirnbauerd}. Secondly, models where stochasticity is
implemented in terms of {\it subgroups} of the ${\rm U}(N)$, e.g. as in
the QCD-Hamiltonian (${\rm SU}(N) \subset {\rm U}(N)$) or the
random Ising Hamiltonian (${\cal Z}_2 \subset {\rm U}(1)$), can be
analysed in terms of suitably adapted versions of the colour-flavour
transformation\cite{Zirnbauere}. Some of these generalisations are
currently under investigation.

We would like to thank A. V. Andreev, S. Hikami, M. Janssen,
V. Kravtsov, P. B. Littlewood, E. M{\"u}ller-Hartmann, A.Tsvelik,
J.J.M. Verbaaschot, and M. R. Zirnbauer for instructive discussions.


\appendix
\section{Choice of the Bosonic Integration Contour}
\label{sec:boson_int}
In the original formulation~\cite{Zirnbauerb} of the colour flavour 
transformation, the integration over the bosonic block of the fields $Z$, 
\[
\int_{{\cal M}_{\rm B}} dZ (\dots)
\]
extended over a manifold ${\cal M}_{\rm B}$ defined by
\begin{equation}
\label{int:MB}
  {\cal M}_{\rm B}=\{(Z,\tilde Z)|\tilde Z = Z^\dagger,\; |{\rm EV
  \,}(Z\tilde Z)|\le 1 \}. 
\end{equation}
Here $|{\rm EV \,}(Z\tilde Z)|\le 1$ is a shorthand
notation for the `modulus of all eigenvalues of the matrix $Z\tilde Z$
smaller than one'.  (As we will be concerned solely with boson
variables throughout, the subscript `BB' in $Z$ has been omitted.)

The Hermitian connection (\ref{int:MB}) between $Z$ and $\tilde Z$
ultimately finds its origin in the interpretation of these fields as parameter
sets of bosonic spin coherent states. (A comprehensive discussion of
the role of Hermiticity in connection with boson spin coherent states
can be found in Perelomov's book~\cite{Perelomov}.) Yet it is {\it not} 
enforced by criteria of mathematical consistency.  In order to encompass the
low-lying modes of the RF-problem it is in fact  necessary to
abandon this connection and to consider pairs $(Z,\tilde Z\equiv
Z^\ddag \not = Z^\dagger )$, where the matrix $Z^\ddag$ has been
defined in Eq.~(\ref{ddag_def}).  More precisely, we need to analytically
continue to a different integration manifold ${\cal N}_{\rm B}$ on
which the connection between $Z$ and $\tilde Z=Z^\ddag$ is still
one-to-one but different from Hermitian adjungation.

Apart from the Hermitian connection between $Z$ and $\tilde Z$, Eq.
(\ref{int:MB}) contains a second condition, viz. $|{\rm EV \,}(Z\tilde
Z)|\le 1$. As opposed to Hermiticity, this criterion is indeed needed
for mathematical reasons. The most straightforward way to see this is
to observe that the BB-block of the logarithmic term appearing in the
colour-flavour transformed action (\ref{Z2}) is given by ${\rm tr\;
  ln\;}(1- Z\tilde Z)$.  The condition $|{\rm EV \,}(Z\tilde Z)|\le 1$
ensures that the cut of the logarithm is not crossed upon integration
over $Z$. In fact, the condition can be relaxed to
\[
{\rm EV \,}(Z\tilde Z) \notin ]1,\infty].
\]
To understand the connection between the integrations over ${\cal
  M}_{\rm B}$ and ${\cal N}_{\rm B}$, respectively, we reconsider the
decomposition
\[
{\rm GL\,}(n)\ni Z=W V, \qquad W\equiv \exp(H) \in {\rm GL\,}(n)/{\rm U\,}(n),\qquad
V=\exp(iH')\in {\rm U\,}(n),
\]
where $H$ and $H'$ are $n$-dimensional Hermitian generators. 
The BB-restricted integration over a function $f(Z,\tilde Z)$ can then be
written as
\[
\int_{{\cal
  M}_{\rm B}} f(Z,\tilde Z) = \int\limits_{ {\rm GL\,}(n)/{\rm
  U\,}(n)\atop  {\rm
  EV}(W)\le 1}  dW \int\limits_{\scriptstyle \rm U(n)} dV f(WV,
  V^{-1}W),
\]
where we have used the fact that $\tilde Z=Z^\dagger = (WV)^\dagger = V^{-1}W$,
and that the eigenvalues of $Z\tilde Z=W^2$ depend only on the
Hermitian factor matrix $W$. The symbols $\int dW$ and $\int dV$
denote an integration over the invariant measures on the coset space
${\rm GL\,}(n)/{\rm U\,}(n)$ and the group ${\rm U\,}(n)$,
respectively.  The basic idea of the analytic continuation is now to
`exchange' the role of the compact ${\rm U\,}(n)$ and the
non-compact ${\rm GL\,}(n)/{\rm U\,}(n)$ submanifolds, respectively.
To this end, notice that both $W$ and $V$ can be unitarily
diagonalised,
\begin{eqnarray*}
    &&W=\exp(H) = U \exp (D) U^\dagger \leadsto dW = dU dD
    J(D),\\
    &&V=\exp(iH') = U' \exp (i D') U^{'\dagger} \leadsto dV = dU' dD'
    J'(D').
\end{eqnarray*}
Here $U,U'\in {\rm U\,}(n)$ are the diagonalizing matrices, $D={\rm
  diag\,}(D_1,\dots,D_n)$ and $D'={\rm diag\,}(D'_1,\dots,D'_n)$ are
the diagonal matrices of real eigenvalues $D_i$ and $D_i'$. Finally,
$J(D)$ and $J'(D')$ are Jacobians resulting from the change of
integration variables and the condition ${\rm EV}(W)\le 1$ translates
to $D_i \le 0$.

We next `Wick rotate' the integration over the eigenvalues $D_i$ from
the real to the imaginary axis. Since ${\rm tr\; ln\,}(1-Z\tilde Z)
= \sum_i \ln (1-e^{2D_i})$, the $D_i$'s can safely be sent from
the negative real to the imaginary axis without crossing
singularities. (After the rotation the eigenvalues of $Z\tilde Z$ are
given by $e^{2iD_i}$. I.e. their modulus does not exceed unity.)  As a
result of this manipulation, the integration over ${\rm GL\,}(n)/{\rm
  U\,}(n)$ becomes one over ${\rm U\,}(n)$ and {\em vice versa}. More
precisely,
\begin{eqnarray*}
&&\int_{{\cal
  M}_{\rm B}} f(Z,\tilde Z) = \int\limits_{ {\rm GL\,}(n)/{\rm
  U\,}(n)\atop  {\rm
  EV}(W)\le 1}  \hspace{-0.5cm} dW \int\limits_{ \rm
  U(n)}\hspace{-0.2cm} dV f(WV, 
  V^{-1}W)\stackrel{\rm Wick}{\longrightarrow}\\
&&\hspace{0.5cm}\longrightarrow \int\limits_{ \rm U(n)}\hspace{-0.2cm}  dW
  \int\limits_{ {\rm GL\,}(n)/{\rm 
  U\,}(n)}\hspace{-0.5cm} dV f(WV,
  V^{-1}W)\equiv \int_{{\cal
  N}_{\rm B}} f(Z,\tilde Z).
\end{eqnarray*}
The last line defines the new integration manifold ${\cal N}_{\rm
  B}$. Notice that, on the new manifold, the connection between $Z$ and
  $\tilde Z$ is no longer provided by Hermitian adjunction. Rather, we
  have
\[
Z=\exp(iH) \exp(H'),\qquad \tilde Z =  \exp(-H')\exp(iH) \equiv
Z^\ddag.
\]
\section{Analytic Continuation of the Energy Arguments}
\label{sec:convergence}
This Appendix is devoted to a discussion of the convergence problems
that arise after the colour-flavour transformation in connection with
the Gaussian integration over the bosonic field variables $\psi_{\rm
  B}$. Since we will be exclusively concerned with the bosonic sector
of the theory, all subscripts `B' will be omitted throughout.

That convergence problems occur follows simply from the fact that the
integration over $Z$ extends over field configurations with finite
imaginary eigenvalues. Thus, it looks like the Gaussian integration
over
\begin{equation}
\label{boson_bil}
e^{i\sum 
\overline\psi_{i} s^{1/2}
 \left(\hat{z} +\sum_{j\in N_i} Z_{ij}\right) s^{1/2}
 \psi_{i}}
\end{equation}
does not converge for all values of $Z_{ij}$ implying that the orders
of integration over $\psi$ and $Z$ cannot be exchanged. 

Below we are going to show that this difficulty can be surmounted by
giving the energy arguments $z_m$ a finite imaginary part. More
precisely, it is necessary to analytically continue from $z_m =
\epsilon_m \pm i\delta$ to $z_m = \epsilon_m \pm 2id$. Since no
singularities ($z_m\in$ real axis) are crossed, the continuation
manipulation is permitted and, after it has been redone on the level
of the subsequent $Z$ integration, does not affect the value of the
functional integral.

To understand that a shift into the deep complex region $|z_m|>2d$
solves the convergence difficulties, we consider bilinear forms like
$\tilde\phi^\dagger Z \phi$ where $Z$ is meant to be the BB-block of
any matrix $Z_{ij}$, $\tilde \phi^\dagger \equiv \psi_i^\dagger
s^{1/2}$ and $\phi \equiv s^{1/2}\psi_{i}$. To formulate the following
estimates, it is useful to represent scalar products like the one
above in terms of a bra-ket notation, e.g.
$\tilde \phi^\dagger Z \phi \equiv \langle \tilde \phi | Z | \phi
\rangle$. Now, for any normalized vector $|n\rangle$, $\langle n | Z
Z^\dagger | n \rangle<1$, a result that straightforwardly follows from
the fact that the matrix $ZZ^\dagger$ is unitarily diagonalisable and
has (see below Eq.~(\ref{Z_domain_BB})) eigenvalues smaller than one. Using
this estimate and setting $|\phi \rangle \equiv \langle \phi |\phi
\rangle^{1/2} |n\rangle$, we next obtain
\begin{eqnarray*}
&&|\psi_{i}^\dagger s^{1/2}Z s^{1/2}\psi_{i}|^2 \equiv |\langle \tilde\phi| Z| \phi\rangle|^2  =\\
&&\qquad 
=\langle \tilde n | 
Z |  n \rangle\langle n | Z^\dagger | \tilde n \rangle\; \langle \tilde
      \phi | \tilde \phi \rangle \; \langle 
      \phi | \phi \rangle  \le\\
&&\qquad \le \sum_m \langle \tilde n | 
Z |  m \rangle\langle m | Z^\dagger | \tilde n \rangle \; \langle \tilde
      \phi | \tilde \phi \rangle \; \langle 
      \phi | \phi \rangle = \\
&&\qquad = \langle \tilde n | 
Z   Z^\dagger | \tilde n \rangle \; \langle \tilde
      \phi | \tilde \phi \rangle \; \langle 
      \phi | \phi \rangle  \le \langle \tilde
      \phi | \tilde \phi \rangle \; \langle 
      \phi | \phi \rangle =  (\psi_{i}^\dagger \psi_{i})^2,
\end{eqnarray*}
where $\{ |m\rangle \}$ is meant to be an extension of $|n\rangle$ to
an orthonormalised basis. This result implies that, 
\[
|{\rm Im\;}\psi_{i}^\dagger
s^{1/2}Z s^{1/2}\psi_{i}| = \psi_{i}^\dagger
\psi_{i}
\] 
and further
\[
|{\rm Im\;}\sum_{j\in N_i} \psi_{i}^\dagger
s^{1/2}Z_{ij} s^{1/2}\psi_{i}| \le 
\sum_{j\in N_i} |{\rm Im\;} \psi_{i}^\dagger
s^{1/2}Z_{ij} s^{1/2}\psi_{i}| \le 
2d\psi_{i}^\dagger
\psi_{i}.
\]
On the other hand, after the analytic continuation of the energy
arguments,
\[
|{\rm Im\;} \psi_{i}^\dagger s^{1/2} \hat z s^{1/2}
\psi_{i}| \ge 2d \psi_{i}^\dagger
\psi_{i}. 
\]
Consequently, the sum of the energy and the $Z$-contribution to the
bosonic bilinear form (\ref{boson_bil}) has a positive imaginary part,
implying that the $\psi$-integration converges.
\section{Gradient Expansion}
\label{sec:gradient}
This Appendix is devoted to the explicit derivation of the three
contributions to the action (\ref{Seff}). To simplify the notation we
introduce the convention that all indices $i,i',\dots \in A$, whilst
$j,j',\dots \in B$. 
\subsection{Fluctuation action, $S_{\rm fl}$:}
Setting $\hat z=0$, ignoring
massive modes, $P=\openone$, and substituting the parameterisation
(\ref{Z_vs_PT}) into (\ref{Z3}) we obtain
\begin{equation}
  \label{Sfl1}
  S_{\rm fl}[T]=N\sum\limits_{i} {\rm str \, ln \,}
\left(\sum_{j\in N_i} T_{ij}\right)+N\sum\limits_{j} {\rm
  str \, ln \,} 
 \left(\sum_{i\in N_j} T^{-1}_{ij}\right).
\end{equation}
We next assume the lattice to be embedded in a continuum, and that
the low energy field configurations $\{ T_{ij} \}$ can be extended to
smooth functions $T$ with continuous support. Under these conditions,
it makes sense to Taylor expand
\begin{equation}
  \label{taylor}
  T_{ij} = T_i + \frac{a}{2} \partial_{i\to j} T_i +
  \frac{a^2}{8}\partial^2_{i\to j} T_i +\dots 
\end{equation}
where $a$ is the lattice spacing, $\partial_{i\to j}$ the
directional derivative along the link $i\to j$, and $T_i$ the function
$T$ evaluated at the coordinate of the lattice point $i$.
Substituting this expression into (\ref{Sfl1}) and taking into account 
the fact that $\sum_{j\in N_i} \partial_{i\to j} T_i =0$, we obtain
\begin{eqnarray}
&&S_{\rm fl}[T] =  N\sum_{i} {\rm str \, ln\,} \left(
2d T_{i} \right) +
N \sum_{j} {\rm str \, ln\,} \left(
2d T_{j}^{-1} \right)+\nonumber \\ 
&&\hspace{0.5cm}+\frac{Na^2}{16d}\left(\sum_{i} 
\sum_{j\in N_i}T_{i}^{-1}
\partial^2_{i\rightarrow j} T_i  +
\sum_{j} \sum_{i\in N_j} T_{j}
\partial^2_{j\rightarrow i} T_j^{-1}\right)+{\cal O}(\partial^4).
\label{Sfl2}
\end{eqnarray}
Up to second order in the number of derivatives (the accuracy of the
present expansion scheme), the first two terms cancel each other:
\begin{eqnarray}
 && \sum_{i} {\rm str \, ln\,} (2d T_{i}) +
\sum_{j} {\rm str \, ln\,}(2d T^{-1}_{j})= \sum_{i} {\rm
  str \, ln\,} (T_{i}) -
\sum_{j} {\rm str \, ln\,}(T_{j})=\nonumber \\
&&\hspace{0.5cm}= \sum_{i}\left( {\rm str \, ln\,} (T_{i}) -
\frac{1}{2d} \sum_{j\in N_i} {\rm str \, ln\,}(T_{j})\right)\simeq\nonumber \\
&&\hspace{0.5cm}\simeq \sum_{i}\left( {\rm str \, ln\,} (T_{i}) -
\frac{1}{2d} \sum_{j\in N_i} {\rm str \, ln\,}
(T_{i}+a\partial_{i\rightarrow j} T_i +
\frac{a^2}{2}\partial^2_{i\rightarrow j} T_i )\right)\simeq\nonumber \\
&&\hspace{0.5cm}\simeq -\frac{1}{2d} \frac{a^2}{2} \sum_{i,j\in N_i}
{\rm str } (-T_i^{-1}(\partial_{i\rightarrow j} T_i)
T_i^{-1}(\partial_{i\rightarrow j} T_i) +
T_i^{-1}\partial^2_{i\rightarrow j} T_i)=\nonumber\\
&&\hspace{0.5cm}=-p\frac{a^2}{4d} \sum_{i,j\in N_i} 
{\rm str } (\partial_{i\rightarrow j}T_i^{-1}
\partial_{i\rightarrow j}T_i +
T_i^{-1}\partial^2_{i\rightarrow j} T_i)=0,
\label{Sfl3}
\end{eqnarray}
where the last equality is based on an `integration by parts'. Using the fact
that $\sum_{j\in N_i} \partial^2_{i\rightarrow j} = 2\partial^2$ is
twice the Laplace operator, and that in the continuum limit $\sum_{i\in
  A} \to \frac{1}{2a^d} \int$ the remaining two terms can be 
brought to the form 
\begin{eqnarray}
&&S_{\rm fl}[T] = \frac{Na^2}{8d}\left(\sum_{i} 
{\rm str }(T_{i}^{-1}
\partial^2 T_i)  +
\sum_{j} {\rm str }(T_{j}
\partial^2 T_j^{-1})\right)\nonumber \\ && 
\qquad \simeq\frac{Na^{2-d}}{16 d}\int {\rm str }\left(T^{-1}\partial^2 T +
    T\partial^2 T^{-1}\right) = -\frac{Na^{2-d}}{8 d}\int {\rm str
      }( \partial
    T^{-1} \partial T ).
\label{Sfl}
\end{eqnarray}
\subsection{Frequency action, $S_{z}$:} 
The frequency action is
straightforwardly derived by expanding the action (\ref{Z3}) to
linear order in $\hat z$. Switching directly to a continuum notation,
we obtain
\begin{eqnarray}
  \label{Sz}
&&S_z[T]=S[T]-S[T]\Big|_{\hat z=0}\stackrel{{\cal
 O}(z)}{\longrightarrow}
 -i\frac{N(2d-1)^{1/2}}{4da^d} \int {\rm str\,} \left(\hat z
 (T+T^{-1})\right).
\end{eqnarray}
\subsection{Integration over massive modes and the action, $S_{\rm
 m}$:} 
In order to derive the contribution $S_{\rm m}$ we proceed in a
standard manner and decompose the full action according to
\begin{equation}
\label{STP_gen}
S[P,T]\to S[X,T] = S[T] + S[X] + S_{\rm int}[T,X],
\end{equation}
where we have introduced generators of the massive modes by
$P_{ij}=\exp(X_{ij})$. In (\ref{STP_gen}), $S[T]$ is the pure
Goldstone action with which we have already dealt, $S[X]$ is purely
massive, and $S_{\rm int}[X,T]$ is the contribution that couples
massive and massless modes. As for $S[X]$, the fact that the mass of
the $X$'s is of ${\cal O}(N)$ (see below) justifies an expansion to
quadratic order in $X$. With $S_{\rm int}[X,T]$ the situation is more
complicated in that we will have to expect contributions ${\cal
  O}(X^n\partial T,X^n(\partial T)^2)$\footnote{That there are no
  terms of ${\cal O}(X^n,T)$ follows from the fact that the $\hat z=0$
  action depends only on bilinears $Z\tilde Z$.} where $n$ may be
arbitrarily large. In order to justify a truncation of this series at
quadratic order, $n=2$, we need to employ $N\gg1$ as an expansion
parameter. At first sight it seems that for $N={\cal O}(1)$, we are
facing a serious problem; Cutting the series is not a controlled
approximation, whilst attempting to explicitly compute contributions
with $n\gg1$ is clearly futile. However, as discussed in the main body
of the text, the situation is not that bad. The structure of the final
result for $S_{\rm m}$ will be fixed by the symmetries of the action.
The worst that may happen for $N={\cal O}(1)$ is a change in the
coupling constants (from the numerical value derived below to a
renormalised one).

Up to an accuracy of ${\cal O}(X^2)$, the action $S_{\rm m}$ is given
by
\begin{equation}
  \label{Sm}
  S_{\rm m}[T]=\left \langle S_{\rm int}^{(2)}[X,T] - \frac{1}{2} (
  S_{\rm int}^{(1)}[X,T])^2 \right\rangle, 
\end{equation}
where $S_{\rm int}^{(n)}[X,T]$ represents the contribution to $S_{\rm
  int}[X,T]$ that is of  $n$th order in $X$ and 
\[
\langle \dots \rangle \equiv \int {\cal D}X e^{-S[X]} (\dots)
,
\]
is the functional average over massive modes. The starting point of
our concrete evaluation of these expressions is the $\hat z=0$ action
 \begin{eqnarray}
\label{SPT}
&&\qquad S[X,T]= 
-N\sum_{\langle i, j
    \rangle}{\rm str \, ln \,} 
(1+x^2e^{2X_{ij}})+\nonumber\\
&&\hspace{2.5cm}+N\sum\limits_{i} {\rm str \, ln \,}
\left(\sum_{j\in N_i} e^{X_{ij}}T_{ij}\right)+N\sum\limits_{j} {\rm
  str \, ln \,} 
 \left(\sum_{i\in N_j}  T^{-1}_{ij} e^{X_{ij}}\right).
\end{eqnarray}
Expanding to second order in $X_{ij}$, we obtain
\begin{eqnarray}
\label{Sm1}
&&   S_{\rm m}[X]=-\frac{N(d-1)}{2d^2}\sum_{\langle i, j
    \rangle}{\rm str\,}(X_{ij}^2)-\nonumber\\
&&\qquad\qquad\qquad-\frac{N}{8d^2}
\sum_{i,jj'\in N_i} {\rm str\,} (X_{ij}X_{ij'})-\frac{N}{8d^2}
\sum_{j,ii'\in N_j} {\rm str\,} (X_{ij}X_{i'j}),\\
\label{Sm12}
&&  S_{\rm m}^{(1)}[X,T]=
\frac{N}{2d}\sum_{\langle i, j\rangle}{\rm str\,}
(A_{ij}  X_{ij})
+\frac{N}{2d}\sum_{\langle i, j\rangle}{\rm str\,}
(X_{ij} B_{ij}),\nonumber\\
&& S_{\rm m}^{(2)}[X,T]=-\frac{N}{4d}\sum_{i} {\rm str \,}
\left(
 \frac{1}{d}\sum_{jj'\in N_i}X_{ij'} A_{ij}X_{ij} 
+\frac{1}{2d}\sum_{jj'\in N_i}X_{ij} A_{ij}
X_{ij'}A_{ij'}
-\sum_{j\in N_i}A_{ij}
  X_{ij}^2\right)-\nonumber\\
&&\hspace{2cm}-\frac{N}{4d}\sum_{j} {\rm str\,}
\left(
\frac{1}{d}\sum_{ii'\in N_j}X_{ij} B_{ij}X_{i'j} 
+\frac{1}{2d}\sum_{ii'\in N_j}B_{ij}X_{ij}
B_{i'j}X_{i'j}
-\sum_{i\in N_j}X_{ij}^2B_{ij}\right).
\end{eqnarray}
Here we have introduced the quantities 
\begin{eqnarray}
  A_{ij} \equiv
T_i^{-1}T_{ij}-\openone &\simeq & \frac{a}{2} T_{i}^{-1}\partial_{i\to
  j} T_i,\nonumber\\
  B_{ij}\equiv T^{-1}_{ij}T_j -\openone &\simeq & \frac{a}{2} T_{j}^{-1}\partial_{i\to
  j} T_j.
\end{eqnarray}

To prepare the perturbative integration we introduce the notation
\begin{equation}
  S_{\rm m}[X]\equiv N
\sum_{
  \langle i, j\rangle\atop
  \langle i', j'\rangle}
{\rm str \,}\left(X_{ij}K^{-1}_{ij,i'j'}X_{i'j'}\right),
\end{equation}
where the kernel $K^{-1}$ is defined through (\ref{Sm1}).  The result
of the integration over $X$ will be essentially determined by the
properties of the inverse $K$. For our purposes, rather than computing
this quantity explicitly, it suffices to realize its two main
features: (a) $K$ is short-ranged (decays exponentially on a scale set
by the level spacing). This follows from the fact that $K^{-1}$ is
massive and `tridiagonal' on the lattice of links: More specifically,
$K^{-1}\sim \hat \Delta + m^{2}$, where $\Delta$ is the lattice
Laplace operator and $m^2$ a mass of ${\cal O}(1)$. (b) $K$ is
symmetric.

Functional expectation values of the contributions $S^{(n)}_{\rm
  int}[X,T]$ can now be computed by means of the contraction rules
\begin{eqnarray}
&&  \langle {\rm str\,}(A  X_{ij}){\rm str\,} (A'  X_{i'j'})
  \rangle_X = \frac{1}{N} K_{ij,i'j'} {\rm str \,}(A 
  A'),\nonumber\\   
&& \langle {\rm str\,}(A X_{ij} A' X_{i'j'})
  \rangle_X = \frac{1}{N} K_{ij,i'j'} {\rm str \,}(A) {\rm str
  \,}(  A').  
\end{eqnarray}
Instead of working out of all possible contractions of the terms
appearing in (\ref{Sm12}) explicitly, it is more efficient to
``pre-exclude'' certain contributions by virtue of their
infrared-irrelevancy or other reasons. Specifically, the contraction
of $(S_{\rm m}^{(1)}[X,T])^2$ gives rise to contributions of the type
$\sim {\rm str}(T_{i}^{-1}\partial_{i\to j} T_i
T_{i'}^{-1}\partial_{i'\to j'} T_i')$. Since the only gradient
operator compatible with the fundamental symmetries of the model is
given by $\sim {\rm str}(\partial T \partial T^{-1})$, contributions
of this type can lead to no more than a renormalization of the
coupling constant in $S_{\rm fl}[T]$. Moreover, the structure of the
kernel $K_{ij,i'j'}$ implies that the coupling constant does not change
  significantly, so that the contribution above can be ignored.

  As for the contraction of $S_{\rm m}^{(2)}[X,T]$ the 1st, 3rd, 4th
  and 6th contribution to that part of the action vanish by
  supersymmetry. The 2nd and 5th term give
\begin{eqnarray*}
&&  S_{\rm m}[T]=\langle S_{\rm int}^{(2)}[X,T]\rangle=
-\frac{a^2}{32d^2}\Bigg(\sum_{i,jj'\in N_i}{\,\rm str\,}(
T_{i}^{-1}\partial_{i\to   j} T_i) 
{\,\rm str \,}( T_{i}^{-1}\partial_{i\to
  j'} T_i)K_{ij,ij'}+\nonumber\\
&&\qquad\qquad
+\sum_{j,ii'\in N_j}{\,\rm str\,}(T_{j}^{-1}\partial_{i\to j} T_j)
{\,\rm str\,}(T_{j}^{-1}\partial_{i'\to j} T_j)K_{ij,i'j}
\Bigg)\simeq\nonumber\\
&&\qquad\qquad\simeq -\frac{a^2}{16d^2}\sum_{i,jj'\in N_i}{\,\rm
    str\,}( T_{i}^{-1}
\partial_{i\to  j} T_i) {\,\rm str \,}( T_{i}^{-1}\partial_{i\to
  j'} T_i)K_{ij,ij'}.
\end{eqnarray*}
It is a straightforward matter to check that, by the symmetry of the
kernel $K$, only the  contributions with $j'=j$ and $j'={\rm o}j$
survive, where ${\rm o}j$ is the site opposite (with respect to $i$) to
$j$. Focusing on the two non-vanishing contributions, we obtain
\begin{eqnarray}
\label{Smf}
  &&  S_{\rm m}[T]\simeq -\frac{a^2C}{16d^2}\sum_{\langle
    i,j \rangle }{\,\rm
    str\,}( T_{i}^{-1}
\partial_{i\to  j} T_i) {\,\rm str \,}( T_{i}^{-1}\partial_{i\to
  j} T_i)\simeq\nonumber\\
&&\qquad\qquad\simeq -\frac{C}{16d^2a^{d-2}}\int {\,\rm
    str\,}( T^{-1}\partial T) {\,\rm str \,}( T^{-1}\partial   T),
\end{eqnarray}
where the constant 
\begin{equation}
  \label{Cdef}
  C=K_{ij,ij}-K_{ij,i{\rm o}j}
\end{equation}
is independent of $i$ by translational symmetry. 
\subsection{Boundary Action, $S_{\rm b}$}
Thus far our derivation of the action has implicitly assumed that the
system we are dealing with is infinitely extended. (E.g. in deriving
(\ref{Sfl}) we have repeatedly integrated by parts with no account for
boundary terms.) For finite size systems, however, the action does
include boundary terms. Even for large systems, these terms must not
be neglected the reason being that some of them are topological and
hence may affect the phenomenology of the system on arbitrary length
scales.

\begin{figure}
\begin{center} 
\epsfig{file=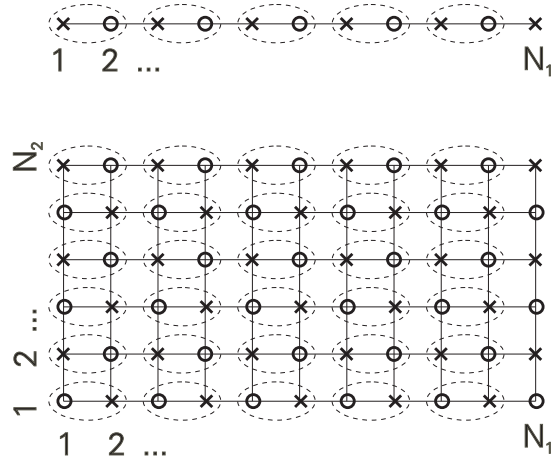,height=6cm}
\end{center}
\caption{\label{fig:lattice} One- and two-dimensional finite size
  lattice with $N_1$=odd sites in $1$-direction and $N_2$=even sites
  in $2$-direction ($2d$ case). The dashed ovals indicate the pairing
  of lattice sites employed in the computation of the boundary
  action.}
\end{figure}

To derive the boundary action, we assume that our (hypercubic) system
has $N_i,i=x,y,\dots$ lattice sites in $x,y,\dots$ direction (c.f.
fig. \ref{fig:lattice}.) As before, the lattice spacing is set by
$a$. As usual with field theories of disordered systems
\cite{Efetovsbook} the constraint that no current flow through the interfaces 
translates to the condition $\partial_\perp T\Big|_{\rm boundary}=0$,
where $\partial_\perp$ is the derivative normal to the boundary. Under
these conditions, the only source for potential boundary contributions
are the first two terms in Eq. (\ref{Sfl2})\footnote{For reasons of
  operator relevancy a boundary operator must not contain more than
  one derivative. Due to current conservation, normal derivatives are
  not permitted. It is a straightforward matter to check that under
  these conditions the first two contributions to (\ref{Sfl2}) are the
  only ones that qualify as potential sources of boundary terms.} As
shown in Eq. (\ref{Sfl3}), these two terms do not contribute to the
bulk action. In order to explore their role at the boundaries, we
imagine that the sites of the lattice are paired to doublets as
indicated in fig. \ref{fig:lattice}

Considering the case of a $1d$-chain first we thus obtain,
\begin{eqnarray}
  \label{Sb1}
  S^{(d=1)}_{\rm b}[T] &=&  N\sum_{i} {\rm str \, ln\,} \left(
2d T_{i} \right) +
N \sum_{j} {\rm str \, ln\,} \left(
2d T_{j}^{-1} \right)=\nonumber\\
&=&N\sum_{i=1,3,\dots,R_1}\left({\rm str \, ln\,} (T_i ) -
 {\rm str \, ln\,} (
 T_{i+1}^{-1} )\right)+N\frac{1-(-)^{N_1}}{2}{\rm str \, ln\,} \left(
T_{N_1} \right) \simeq\nonumber\\
&\simeq&-N\frac{1}{2}\int_0^L dx \partial_x {\rm str \, ln\,} (T(x) )+N\frac{1-(-)^{N_1}}{2}{\rm str \, ln\,} \left(
T((N_1-1)a) \right) =\nonumber\\
&=&\frac{N}{2}\left({\rm str \, ln\,} (T(0)) -(-)^{N_1}{\rm str \,
    ln\,} (T(L))\right), 
\end{eqnarray}
where $R_1=N_1-1$ for $N_1$ even and $R_1=N_1-2$ for $N_1$ odd. That
scheme can straightforwardly be generalized to higher
dimensions. E.g. in $d=2$, we obtain
\begin{eqnarray*}
  S^{(d=2)}_{\rm b}[T] &=&\frac{N}{4}\Big(
(-)^{N_1+N_2}{\rm str \, ln\,} (T(L_1,L_2))
-(-)^{N_1}{\rm str \, ln\,} (T(L_1,0))-\\
&&\qquad-(-)^{N_2}{\rm str \, ln\,} (T(0,L_2))+
{\rm str \, ln\,} (T(0,0))\Big).
\end{eqnarray*}
The obvious generalisation to arbitrary dimensions is given by
\begin{equation}
  \label{Sb_app}
  S_{\rm b}[T]=\frac{N}{2^d}\sum_{s_i=0,1} 
(-)^{\sum_{i=1}^d (N_i+1)s_i} {\rm str \, ln\,} (T(s_1 L_1,\dots,s_dL_d)).
\end{equation}
Summing the contributions (\ref{Sfl}), (\ref{Sz}), 
(\ref{Smf}) and (\ref{Sb_app}) we obtain the full continuum action 
Eq.~(\ref{Seff}).
\section{The Zero Mode Integral}
\label{sec:0-mode}
The computation of the integral (\ref{DoS1}) can be most readily performed
by employing the `polar' parameterisation of the supermatrices,
\begin{eqnarray}
\label{polar}
  &&T=k a k^{-1},\nonumber\\
&&\qquad k=\exp\left(
  \begin{array}{cc}
&\eta\\
\nu&
  \end{array}\right), \qquad a=\exp\left(
  \begin{array}{cc}
x&\\
&iy
  \end{array}\right),
\end{eqnarray}
where $\eta$ and $\mu$ are Grassmann variables and $x,y\in {\cal R}$.
The invariant measure associated with (\ref{polar}) is given by
\begin{equation}
  \label{measure}
  \int dT f(T) \to \int d\eta d\mu \int_0^\infty dx \int_0^{2\pi} dy
  \frac{1}{\sinh^2(\frac{x-iy}{2})} f(\eta,\mu,x,y).
\end{equation}

With this definition, the DoS takes the form
\begin{eqnarray*}
\rho(\epsilon)=\rho_0\left[1+{1\over 8}{\rm Re}\int_{-\infty}^\infty dx
\int_0^{2\pi} dy f(x,y){\cosh x-\cos y\over\sinh^2((x-iy)/2)} 
e^{is_+(\cosh x-\cos y)}\right]
\end{eqnarray*}
where $s_+=\pi\rho_0\epsilon_+$, and
\begin{eqnarray*}
f(x,y)=\cases{e^{-x+iy} &$N_{\rm sites}~{\rm odd}$\cr
1 &$N_{\rm sites}~{\rm even}$.\cr}
\end{eqnarray*}
For a lattice with an even number of sites, one can show
\begin{eqnarray*}
\partial_s\rho=\rho_0{1\over 2}{\rm Re}i\int_0^\infty dx
\int_0^{2\pi} dy (\cosh x\cos y-1)e^{is_+(\cosh x-\cos y)}
\end{eqnarray*}
Then, making use of the identities,
\begin{eqnarray*}
{\rm J}_n(s)={1\over 2\pi}\int_0^{2\pi} e^{i(s\sin y-ny)}={2\over\pi}
\int_0^\infty dx \cosh(nx)\sin(s\cosh x-\pi n/2),
\end{eqnarray*}
we find $\partial_s\rho=\rho_0\pi^2(J_0^2(s)-J_1^2(s))/2$. Finally, making 
use of the identity
\begin{eqnarray*}
\partial_s\left[s\left(J_{n+1}^2(s)+J_n^2(s)\right)\right]=(2n+1)\left(
J_n^2(s)-J_{n+1}^2(s)\right),
\end{eqnarray*}
one obtains the standard result from ChRMT shown in Eq.~(\ref{rho}).

For the lattice with an odd number of sites, we have the identity
\begin{eqnarray*}
{1\over 2}{\rm Re}\left({e^{-x+iy}\over \sinh^2((x-iy)/2)}+{e^{x+iy}\over 
\sinh^2((x+iy)/2)}\right)=2\left[1+{\cosh x\cos y-1\over (\cosh x-\cos y)^2}
\right]
\end{eqnarray*}
As a result, the expression for the DoS separates into two contributions
$\rho(\epsilon)=\rho_{\rm even}(\epsilon)+\delta\rho(\epsilon)$. The first 
reproduces the expression for the DoS with an even number of lattice 
sites~(\ref{rho}) while the second generates the correction,
\begin{eqnarray*}
\delta\rho(\epsilon)={\rho_0\over 2}{\rm Re}\int_0^\infty dx
\int_0^{2\pi} dy (\cosh x-\cos y)e^{is_+(\cosh x-\cos y)}
\end{eqnarray*}
Making use of the identities above we find $\delta\rho(s)=-\pi^2 J_0(s)J_1(s)$
which, taken together gives the expression for the DoS shown in 
Eq.~(\ref{rho2}) for the case $N=1$. For arbitrary $N$, a calculation~\cite{v4}
analogous to the one above obtains the more general formula~(\ref{rho2}).

\end{document}